\documentclass[a4paper,12pt]{article}

\usepackage[labelfont=bf,justification=justified,margin=2cm]{caption}
\usepackage[square,compress,numbers,comma]{natbib}

\usepackage{upgreek}
\usepackage{lipsum}
\usepackage{graphicx,dblfloatfix}
\usepackage{subfig}

\usepackage{hyphenat}
\graphicspath{\home}
\usepackage{wrapfig}

\usepackage{amsmath}

\numberwithin{equation}{section}

\usepackage{amssymb}
\usepackage{cancel}
\usepackage{scalerel}
\usepackage{pifont}
\usepackage{nicefrac}

\usepackage{vwcol}

\usepackage{multicol}
\usepackage{siunitx}
\usepackage{xcolor} 
\usepackage[]{hyperref}
\hypersetup{pdfborder={0 0 0},
	colorlinks=true,
	hyperindex=true,	
	linkcolor = blue,
	citecolor=blue,
	urlcolor=blue}

\makeatletter
\newcommand\notsosmall{\@setfontsize\notsosmall{9}{9.2}}
\makeatother

\newcommand{\stack}[2]{\stackrel{\scriptsize \mbox{#1}}{#2}}

\newcommand\rhor{\rho_{_2}}
\newcommand\refr{\rho^{\mathrm{eff}}_{2}}
\newcommand\rer{\rho_{_{2,\mathrm{ext}}}}
\newcommand\tre{t_{\mathrm{2}}}
\newcommand\rphir{\varphi_{2}}

\newcommand\rhol{\rho_{_1}}
\newcommand\refl{\rho^{\mathrm{eff}}_{1}}
\newcommand\rel{\rho_{_{1,\mathrm{ext}}}}
\newcommand\tle{t_{\mathrm{1}}}
\newcommand\rphil{\varphi_{1}}

\newcommand\rhoi{\rho_{_j}}
\newcommand\refi{\rho^{\mathrm{eff}}_{j}}
\newcommand\rei{\rho_{_{j,\mathrm{ext}}}}
\newcommand\tie{t_{j}}
\newcommand\rphii{\varphi_{j}}

\newcommand\kr{\kappa_{2}}
\newcommand\kl{\kappa_{1}}
\newcommand\ki{\kappa_{j}}

\newcommand\wfb{\omega_{_\mathrm{FB}}}
\newcommand\wref{\omega_{\mathrm{ref}}}
\newcommand\tcav{t_{\mathrm{cav}}}

\newcommand\phasedr{\phi_{2}}
\newcommand\phasedl{\phi_{1}}

\newcommand\sqsl{\mathcal{S}_{1}^{+}}
\newcommand\sqsr{\mathcal{S}_{2}^{-}}
\newcommand\sqsi{\mathcal{S}_{j}^{\pm}}

\newcommand\phitl{\phi^{+}_{\tle}}
\newcommand\phitr{\phi^{-}_{\tre}}
\newcommand\phiti{\phi^{\pm}_{\tie}}

\newcommand\Phitl{\Phi^{+}_{1}}
\newcommand\Phitr{\Phi^{-}_{2}}

\newcommand\phim{\phi_{\mathrm{m}_{j}}}

\newcommand\I{S}
\newcommand\meanI{S_{0}}
\newcommand\pddev{\int_{-\infty}^{\infty} e^{i\freq' t} S_{\mathrm{0p}} (\freq')  \mathrm{d} \freq'}
\newcommand\pd{\I_\Delta}
\newcommand\dpd{i \freq' \pd}
\newcommand\hatpd{S_{\mathrm{0p}}}

\newcommand\phipdev{\int_{-\infty}^{\infty} e^{i\freq' t}\phi_{\mathrm{0p}} (\freq')  \mathrm{d}\freq'}
\newcommand\phip{\phi_\Delta}
\newcommand\hatphip{\phi_{\mathrm{0p}}}
\newcommand\dotphip{i \freq' \phip}

\newcommand\nopdev{ \int_{-\infty}^{\infty}e^{i\freq' t} N_{\mathrm{0p}} (\freq')  \mathrm{d}\freq'}
\newcommand\nop{N_\Delta}
\newcommand\meanN{N_0}
\newcommand\dotnop{i \freq' \nop}
\newcommand\hatnop{N_{\mathrm{0p}}}

\newcommand\FI{F_\I}
\newcommand\hatFI{\widehat{F}_{\I}}

\newcommand\Fphi{F_\phi}
\newcommand\hatFphi{\widehat{F}_\phi}

\newcommand\FN{F_N}
\newcommand\hatFN{\widehat{F}_N}

\newcommand\phased{\phi_{{j}}}
\newcommand\kc{\kappa_{j}^\mathrm{c}}
\newcommand\kcr{\kappa_{\mathrm{2}}^\mathrm{c}}
\newcommand\kcl{\kappa_{\mathrm{1}}^\mathrm{c}}

\newcommand\ks{\kappa_{j}^\mathrm{s}}
\newcommand\ksr{\kappa_{\mathrm{2}}^\mathrm{s}}
\newcommand\ksl{\kappa_{\mathrm{1}}^\mathrm{s}}

\newcommand\ah{\alpha_{_\mathrm{H}}}
\newcommand\diffgain{a}
\newcommand\again{a_g}
\newcommand\taue{t_{\mathrm{ext}}}
\newcommand\dif{\mathrm{d}}
\newcommand\gamman{\tau_{\mathrm{e}}^{-1}}
\newcommand\gammansq{\tau_{\mathrm{e}}^{-2}}
\newcommand\gammai{\zeta_{_\I}}

\newcommand\freq{\Omega}
\newcommand\deltam{B_{\phi}}
\newcommand\AN{A_{N}}
\newcommand\AI{A_{\I}}
\newcommand\Aphi{A_{\phi}}
\newcommand\tcoh{t_{\mathrm{coh}}}
\newcommand\rrec{N \tau_\mathrm{sp}^{-1}} 

\newcommand\Rbim{ R_{\mathrm{sp}} } 
\newcommand\R{R_{\mathrm{sp}}} 
\newcommand\rrecmean{N_{0} \tau_{\mathrm{sp}}^{-1}}
\newcommand\rrecN{\tau_{\mathrm{sp}}^{-1}}
\newcommand\gain{g_{_{\mathrm{FB}}}}

\newcommand\vg{ v_{\mathrm{g}} }
\newcommand\gvg{\Gamma \vg}
\newcommand\go{G_0}

\newcommand\G{\Gamma \gain \vg}

\newcommand\tph{\tau_{\mathrm{ph}}^{-1}}

\newcommand\GnG{\again  \go }

\newcommand\dnn{D_{NN}}
\newcommand\dii{D_{\I \I}}
\newcommand\dpp{D_{\phi \phi}}
\newcommand\din{D_{\I N}}

\newcommand\deltafour{\Delta_{4}}
\newcommand\deltatwo{\Delta_{2}}
\newcommand\deltazero{\Delta_{0}}

\newcommand\PSD{S_{f}^{(1)}}

\title{Dynamics and stability conditions of semiconductor lasers under external optical feedback from both sides of the laser cavity}

\author{Mónica Far Brusatori $^{1}$* and Nicolas Volet $^{1}$ \\ *mfar@ece.au.dk \\ $^{1}$ Department of Electrical and Computer Engineering\\ Aarhus University, Aarhus, Denmark}

\date{}

\begin{document}
	\maketitle
\abstract{To increase the spectral efficiency of coherent communication systems, lasers with ever-narrower linewidths are required as they enable higher-order modulation formats with lower bit-error rates. In particular, semiconductor lasers are a key component due to their compactness, low power consumption, and potential for mass production. In field-testing scenarios their output is coupled to a fiber, making them susceptible to external optical feedback (EOF). This has a detrimental effect on its stability, thus it is traditionally countered by employing, for example, optical isolators and angled output waveguides. In this work, EOF is explored in a novel way with the aim to reduce and stabilize the laser linewidth.
EOF has been traditionally studied in the case where it is applied to only one side of the laser cavity. In contrast, this work gives a generalization to the case of feedback on both sides. It is implemented using photonic components available via generic foundry platforms, thus creating a path towards devices with high technology-readiness level. Numerical results shows an improvement in performance of the double-feedback case with respect to the single-feedback case. In particularly, by appropriately selecting the phase of the feedback from both sides, a broad stability regime is discovered. This work paves the way towards low-cost, integrated and stable narrow-linewidth integrated lasers.}

\setlength{\jot}{15pt}

\section{Introduction}

The effect of external optical feedback (EOF) on diode laser dynamics has been extensively studied for the past half century~\cite{Paoli1970,Kanada_1979,Nilsson1981,Kikuchi_1982,Patzak_1983-1,Spano_1984, Tromborg_1987,Petermann_1995,Yousefi_1999,Sattar_2015, van_Schaijk_2018_theory,Happach_2020_a,Ahmed_2021}. EOF has been proven to affect laser performance, showing regimes that can aid in linewidth reduction \cite{Agrawal1984, Olesen1986, Schunk_1988,Patzak_1983}, as well as others responsible for highly unstable behavior, from mode hopping to the case of coherence collapse ~\cite{Tkach1986_1,Lenstra_1985, Henry_1986,Ebisawa_2020,Locquet_2020,Gomez_2020}. Methods to improve laser stability thus need to take EOF into account, as even weak feedback can be detrimental. A traditional approach to mitigate its effects is to include an off-chip isolator at the laser output. Yet, this component negatively impacts the dimensions of packaged devices as well as fabrication times and costs. As such, research is ongoing to develop an integrated solution that can minimize the negative effects of EOF.
Efforts include adjusting the feedback phase to tune into line-narrowing regimes~\cite{Zhao2018,Aoyama2018_2}, using unidirectional phase modulators~\cite{Doerr_2011,van_Schaijk_2018}, reducing the linewidth enhancement factor \textit{e.g.} using quantum dots~\cite{Septon2019,Duan_2019,Yamasaki_2021}, employing electromagnetic effects~\cite{Huang2018,Yan_2020}, harnessing the mode propagation properties of ring lasers~\cite{Lenstra_2019,Khoder_2016}, or the extended cavity approach~\cite{Komljenovic2017, Kasai2018,Morton2018,Xiang2019}.

The established line of thought relies on the key assumption that feedback is introduced from only one side of the laser cavity. Current integration technologies make this assumption obsolete, as they allow for arbitrarily complex design geometries with a variety of functionalities, such as tunability and modulation, while maintaining narrow-linewidth performance~\cite{Barton2003, DHuang2019,Wang2017, Lin2018,Li_2021}. Consequently, this work aims to extend the theoretical foundations of EOF to the case of feedback coupling into the laser cavity from both sides. This system is studied to obtain and analyze its dynamic rate equations.
Furthermore, the frequency noise power spectral density and subsequently the intrinsinc linewidth's dependence on feedback is computed. The Lang-Kobayashi approach is used~\cite{Lang1980}, where an additional term to account for the extra feedback term is included. In a similar fashion, to obtain analytical solutions both small-signal and weak feedback conditions are assumed. The obtained equations are then numerically solved. 
Results show the existence of a feedback-insensitive regime achieved by tuning the feedback parameters. This regime is feasible using mature components of photonic integrated circuits available in generic foundry platforms~\cite{Smit_2014, Augustin2018}, thus creating a path towards devices with high technology-readiness level while maintaining low cost and size.




\section{Rate equations model}
This section includes a comprehensive derivation the dynamic equations of a laser cavity with EOF under the revised assumption that feedback couples into the laser cavity from both sides. Starting from the Lang-Kobayashi model~\cite{Lang1980}, the lasing frequency and threshold gain shifts, as well as the change of intrinsic linewidth due to feedback are obtained. 

\begin{figure}[]
	\begin{center}
		\includegraphics[width=0.9\linewidth]{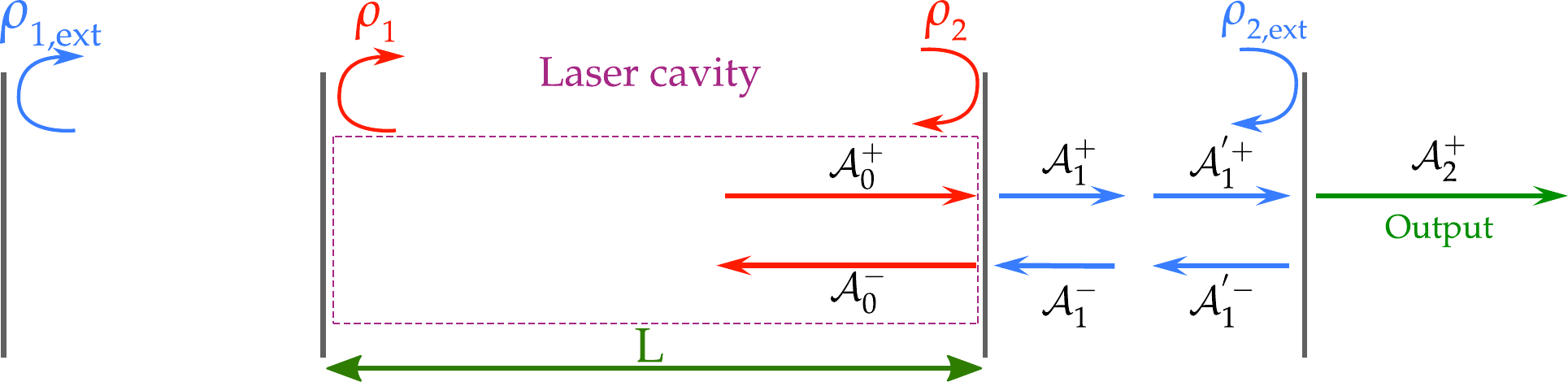}
		\caption{Schematic of a laser cavity affected by external optical feedback from both sides.}
		\label{fig:lasercav}
	\end{center}
\end{figure}

In contrast with previous literature, this work proposes a revised laser system, as shown in \mbox{Fig.~\ref{fig:lasercav}}.
The laser cavity of length $L$ is delimited by two mirrors with complex reflection coefficients $\rhol$ and $\rhor$ respectively. 
Assuming two interfaces at each side of the main cavity, two additional back-reflections ($\rel$ and $\rer$) have to be considered, which can be accounted for by calculating effective reflection coefficients. 
To simplify the problem, this analysis does not take into account multiple reflections in the external cavities.
This is justified for:
\begin{equation}
\label{eq:weak_feedback}
|\rei \rhoi|\ll 1,
\end{equation}
%
%
which includes both weak feedback 
\mbox{$|\rei| \ll |\rhoi|$} and strong feedback \mbox{$|\rei| \gg |\rhoi|$}.

\noindent The following parameters are introduced: 
\begin{subequations} 
	\label{eq:defforgain}
	\setlength{\abovedisplayskip}{0pt}
	\setlength{\belowdisplayskip}{0pt}
	\setlength{\columnsep}{0em}
	\begin{multicols}{2}
		\begin{alignat}{2}
		\ki  & \equiv \frac{1-\left|\rhoi\right|^2}{\rhoi} \frac{\rei}{\tcav}
		\label{eq:kappa} \\
		\phased  &\equiv \omega_{\mathrm{FB}}\tie  + \phim \label{eq:phased}\\	
		\delta \omega &\equiv \wfb - \wref,	\label{eq:deltaomega}		
		\end{alignat}
		
		\begin{align}
		\tcav & = 2 L / \vg , \label{eq:tcav}\\	
		\beta & \equiv n \omega / c \;, \label{eq:beta}
		\end{align}		
	\end{multicols}
\end{subequations}
\noindent  
with $j = 1,2$, where $\ki$ is the coupling coefficient;
$\tcav$ is the cavity roundtrip time, with the group velocity $\vg$;
$\phased$ is the phase delay due to the external cavities determined by the external roundtrip time $\tie$, the lasing frequency in the presence of feedback $\wfb$, and a phase shift at the external mirrors $\phim$;
$\omega_{\mathrm{ref}}$ is the free running laser frequency;
$\beta$ is the propagation constant;
$n$ the effective refractive index of the lasing mode;
and $c$ the speed of light in vacuum.
The first step is extracting the lasing conditions of the proposed laser system.

\subsection{Lasing conditions}
\label{sec:ampphase}
By analyzing $\mathcal{A}$, the slowly-varying amplitude of the complex electric field drawn in \mbox{Fig.~\ref{fig:lasercav}}, the effective reflection coefficients $\left(\refi\right)$ are obtained in \mbox{Appendix~\ref{ap:rhoef}}:
\begin{alignat}{2}
\frac{\refi}{ \rhoi} &  \stack{(\ref{eq:rhoFinal_app})}{\approx} 1 + \ki \tcav e^{\pm i\phased}
= 1 + \ki \tcav  \cos(\phased) 
\pm i \ki \tcav \sin(\phased), \label{eq:rhoFinal}
\end{alignat}
where the plus and minus signs corresponds to $j = 1$ (left mirror) and $j = 2$ (right mirror),  respectively. 
The additional reflection influences the lasing condition, as shown in \mbox{Appendix~\ref{ap:conditions}}:
\begin{alignat}{2}
1 & 
\stack{(\ref{eq:ampexplicit_app})}{\approx} \rhor \rhol \left[1 + \kr \tcav \cos(\phasedr)\right] \left[1 + \kl \tcav \cos(\phasedl)\right] e^{\left(\Gamma \gain-\alpha \right) L}  \label{eq:ampexplicit}\\
2\pi m   &
\stack{(\ref{eq:phaseexplicit_app})}{\approx}  2\beta L + \kr \tcav \sin(\phasedr) - \kl \tcav \sin(\phasedl)\quad , \quad m\in \mathbb{Z}.\label{eq:phaseexplicit}
\end{alignat}
where $\Gamma$ is the confinement factor, 
$\alpha$ is the attenuation coefficient,
and
$\gain$ is the threshold gain coefficient with feedback.
The interplay between the feedback parameters $\kl$, $\kr$, $\phasedl$ and $\phasedr$ determines the dynamics and stability of the system. 
The calculations to obtain the threshold gain reduction and lasing  frequency shift due to EOF are shown in \mbox{Appendix~\ref{ap:threshfreqshift}}. 
The following definitions are convenient: %
\begin{subequations}
	\setlength{\abovedisplayskip}{0pt}
	\setlength{\belowdisplayskip}{0pt}
	\setlength{\columnsep}{0em}
	\begin{multicols}{2}
		\begin{alignat}{2}
		G_{_\mathrm{FB}} & \equiv \G \label{eq:bigG} \\
		G_{\mathrm{th}} & \equiv \Gamma g_{\mathrm{th}} \vg  \label{eq:bigGth} \\
		\delta G & \equiv G_{_\mathrm{FB}}  - \tph
		\end{alignat}		
		\setlength{\abovedisplayskip}{0pt}
		\setlength{\belowdisplayskip}{0pt}
		\setlength{\columnsep}{0em}		
		\begin{alignat}{2}
		\gamma_{_{\mathrm{H}}} & \equiv \sqrt{1+\ah^2} \label{eq:gammaagr}\\ 
		\theta_{_{\mathrm{H}}} & \equiv \arctan \left(\ah\right) \; , \label{eq:alphah}
		\end{alignat} 
	\end{multicols}
\end{subequations}
\noindent where 
$ \tau_{\mathrm{ph}} $ is the photon decay time, which accounts for cavity and mirror losses;
$g_{\mathrm{th}}$ is the threshold gain without feedback;
and $\ah$ is the linewidth enhancement factor~\cite{Henry1982}. Thus, from \mbox{Appendix~\ref{ap:threshfreqshift}}, it is possible to obtain:
\begin{subequations}
	\label{eq:gainfreqshift}
	\begin{align} 
	\delta G \equiv G_{_\mathrm{FB}} - G_{\mathrm{th}} & \stack{(\ref{eq:gainfromamp_app})}{\approx} - 2\kr \cos(\phasedr) - 2\kl \cos(\phasedl)  \label{eq:gainfromamp}
	\\
	\delta \omega &
	\stack{(\ref{eq:omegafeedback_app})}{\approx} \hspace{-2mm}
	- \gamma_{_\mathrm{H}} \left[ \kr \sin (\phasedr + \theta_{_{\mathrm{H}}}) + \kl \sin (\theta_{_{\mathrm{H}}} - \phasedl)\right] ,\label{eq:omegafeedback}
	\end{align}		
\end{subequations}
\noindent where both $\phased$ and $\delta \omega$ depend on the lasing frequency. The relation between the right-hand terms determines the shift in threshold gain and lasing frequency with feedback. Given the transcendental form of \mbox{Eq.~(\ref{eq:omegafeedback})}, a numerical analysis under different feedback conditions is studied in \mbox{Sec.~\ref{sec:sims}}. Nevertheless, an analytical solution for lasing  frequency stability can be found for the condition: 
\begin{equation}
\delta \omega = 0 \Rightarrow \wfb = \wref . \label{eq:stablefreq}
\end{equation}
Under this condition, \mbox{Eq.~(\ref{eq:omegafeedback})} can be rewritten as:
\begin{equation}
	\kr \sin (\phasedr + \theta_{_{\mathrm{H}}}) \stack{(\ref{eq:stablefreq})}{=} - \kl \sin (\theta_{_{\mathrm{H}}} - \phasedl) , \label{eq:freq_stable_gen}
\end{equation}
which is determined by the feedback parameters $\ki$ and $\phased$, the latter being dependent on the time delay $\tie$ as well as the lasing frequency. Finding a stable solution that does not depend on the lasing frequency is of particular interest, as it can be advantageous for tunable laser and their numerous applications. With the following assumption:
\begin{equation}
\kr = \kl= \kappa \label{eq:kappa_assumption},
\end{equation}
\mbox{Eq.~(\ref{eq:freq_stable_gen})} can be rewritten as:
\begin{align}
\sin (\phasedr + \theta_{_{\mathrm{H}}}) & = - \sin (\theta_{_{\mathrm{H}}} - \phasedl) \stack{(\ref{eq:parity})}{=}  \sin (\phasedl - \theta_{_{\mathrm{H}}}) \nonumber\\
\Rightarrow \phasedr + \theta_{_{\mathrm{H}}} & =  \phasedl - \theta_{_{\mathrm{H}}} + 2 m \pi .
\end{align}
Without loss of generality, the parameter $m$ is set to $m=0$, thus:
\begin{align}
 2\theta_{_{\mathrm{H}}} & =  \phasedl - \phasedr \stack{(\ref{eq:phased})}{=} \omega_{\mathrm{FB}}(\tle-\tre)  + (\phi_{\mathrm{m}_{1}}-\phi_{\mathrm{m}_{2}}) \nonumber\\
 \Rightarrow 2\theta_{_{\mathrm{H}}} - (\phi_{\mathrm{m}_{1}}-\phi_{\mathrm{m}_{2}}) &= \omega_{\mathrm{FB}}(\tle-\tre) \label{eq:arctan_condition}.
\end{align}
Choosing an equal time delay (\textit{i.e.} length) in both external cavities:
\begin{equation}
\tre = \tle = \taue \label{eq:text_assumption},
\end{equation}
sets the left hand term of \mbox{Eq.~(\ref{eq:arctan_condition})} equal to zero, so that:
\begin{align}
\Delta \phi_{\mathrm{m}} &\equiv \phi_{\mathrm{m}_{1}}-\phi_{\mathrm{m}_{2}} \\
2\theta_{_{\mathrm{H}}} & 
= \Delta \phi_{\mathrm{m}}. \label{eq:phim_condition}
\end{align}
This result shows that by tuning the phase in the external cavities, so that \mbox{condition~(\ref{eq:phim_condition})} is met, it is possible to obtain a feedback-insensitive lasing frequency.
An active method to tune the phase is however required as $\ah$ is dependent on laser parameters, such as carrier density and wavelength~\cite{Osinski_1987}. 
This can be managed by \textit{e.g.} phase shifters, which are mature and widely used components that can be included on-chip in a laser.

The shown stable solution thus requires meeting the \mbox{conditions~(\ref{eq:kappa_assumption})},~(\ref{eq:text_assumption}) \mbox{and~(\ref{eq:phim_condition})}, which constrain the feedback parameters of one side of the cavity with respect to those of the other side, but do not restrict their absolute value. 
Nevertheless, if \mbox{conditions~(\ref{eq:kappa_assumption})} \mbox{and (\ref{eq:text_assumption})} are not met, solutions for stable performance become frequency dependent. This case would thus only be satisfied for certain lasing frequency values for a given set of feedback parameters, which can potentially yield unstable solutions for other frequencies.

Finally, looking at the (\mbox{Eq.~\ref{eq:gainfromamp}}), under \mbox{conditions (\ref{eq:kappa_assumption})} \mbox{and~(\ref{eq:text_assumption})} \mbox{Eq. (\ref{eq:gainfromamp})} becomes:
\begin{align}
	G_{_\mathrm{FB}} - G_{\mathrm{th}} & =  - 2\kappa \left[\cos(\wref \taue + \phi_{\mathrm{m}_1}) + \cos(\wref \taue + \phi_{\mathrm{m}_2})) \right],
\end{align}
which becomes zero if:
\begin{align}
\cos(\wref \taue + \phi_{\mathrm{m}_1}) & = -\cos(\wref \taue + \phi_{\mathrm{m}_2}) \nonumber\\
\cos(\wref \taue + \phi_{\mathrm{m}_1}) & = \cos(\wref \taue + \phi_{\mathrm{m}_2}- \pi) \nonumber\\
 \Rightarrow \wref \taue + \phi_{\mathrm{m}_1} &= \wref \taue + \phi_{\mathrm{m}_2}- \pi + 2 m \pi \quad m \in \mathcal{N} \nonumber \\
 \stack{m=0}{\Rightarrow}\Delta \phi_{\mathrm{m}} = \pi. \label{eq:gainpi}
\end{align}
This condition, while different than \mbox{condition (\ref{eq:phim_condition})}, also yields stability regardless of lasing frequency in this case for the threshold gain. Both cases are studied numerically in \mbox{Sec.~\ref{sec:sims}}.

\subsection{Rate equations}
In order to obtain the frequency noise (FN) power spectral density (PSD), and from it the laser linewidth, the laser rate equations for the intensity and phase, as well as one for the carrier density need to be studied. The former two can be  extracted from the dynamic equations for the field inside the laser cavity, following the Lang-Kobayashi~\cite{Lang1980} approach. Its full derivation is shown in \mbox{Appendix~\ref{sec:aprates}}. Furthermore, Langevin noise terms are included to account for shot noise fluctuations.
The following definitions are useful to simplify notation:
%
%
%
\begin{subequations}
	\label{eq:fielddefinitions}
	\setlength{\abovedisplayskip}{0pt}
	\setlength{\belowdisplayskip}{0pt}
	\setlength{\columnsep}{0em}	
	\begin{multicols}{2}
		\begin{align}
		\mathcal{A}(t) & = \sqrt{\I (t)} e^{-i\phi(t)},  \label{eq:amplitude_decomp} \\
		\sqsi & \equiv \ki \sqrt{S(t \pm \tie)} \label{eq:sqs} \\
		\phiti & \equiv \phi(t \pm \tie)\label{eq:phit} 
		\end{align} 
		\setlength{\abovedisplayskip}{0pt}
		\setlength{\belowdisplayskip}{0pt}
		\setlength{\columnsep}{0em}	
    	\begin{alignat}{2}
		\Delta \Phitl &\stack{(\ref{eq:phit})}{=} \phi(t)-\phitl - \phasedl \label{eq:deltaphitl}\\
		\Delta \Phitr &\stack{(\ref{eq:phit})}{=} \phi(t)-\phitr + \phasedr \label{eq:deltaphitr} 
		\end{alignat}		
	\end{multicols}
\end{subequations}
%
%
%
%
%
%
\noindent with $j = 1,2$ relating to the EOF components from the right and left respectively. Assuming that the field amplitude $\mathcal{A}$ is slowly varying, where $\I $ is the photon number inside the laser cavity and $\phi$ is the phase of the field, the rate equations of the system can be written as:
\begin{subequations}
	\label{eq:rateposta}
	\begin{alignat}{2}
	\dot{\I} & \stack{(\ref{eq:finalPhotonRate})}{=} \I  \Delta G  
	+ 2 \sqsr \sqrt{\I }\cos(\Delta \Phitr) + 2 \sqsl \sqrt{\I}\cos(\Delta \Phitl)
	+ \Rbim  + F_{\I} \\
	\dot{\phi} & \stack{(\ref{eq:finalPhaseRate})}{=} 
	 \frac{\ah \Delta G}{2} - \delta \omega 
	- \frac{\sqsr}{\sqrt{\I}} \sin\left(\Delta \Phitr \right)
	- \frac{\sqsl}{\sqrt{\I}} \sin\left(\Delta \Phitl \right)
	+F_{\phi}\\
	\dot{N}& \stack{\cite{Agrawal1984}}{=} I - G \I(t)- \rrec +F_N,
	\end{alignat}
\end{subequations}
where $I$ is the effective rate of injected current (in electrons), $\tau_\mathrm{sp}$ is the carrier lifetime, and $\Rbim$ is the spontaneous recombination rate. 
The Langevin noise sources $F_{\I}(t)$, $F_{\phi}(t)$ and $F_{N}(t)$ satisfy~\cite{laxV}:
%
%
%
\begin{subequations}
	\begin{alignat}{2}
	\langle F_{i}(t)\rangle & = 0 \label{eq:langaverage}\\
	\langle F_{i}(t_1) F_{j}(t_2)\rangle & = 2D_{ij} \delta(t_1-t_2) \; \; \; \text{ with }\; i,j = \I ,\, \phi \, \mathrm{ or } \, N , \label{eq:langcross} 
	\end{alignat}
	\label{eq:langeving}
\end{subequations}
where:
\begin{equation}
\dii  = \R  \I \quad ; \quad \dpp  = \frac{\R }{4\I} \quad ; \quad
\dnn  =\R \I + N \tau_\mathrm{sp}^{-1} \quad ; \quad \din  =  - \R  \I , \label{eq:diffusion}
\end{equation}
are standard diffusion coefficients.
Using the Fourier transform:
\begin{equation}
\hat{f}(\freq) \equiv \int _{-\infty }^{\infty }f(t)\ e^{- i \freq t}\,\dif  t, \label{eq:fourier}
\end{equation}
\mbox{Eq.~(\ref{eq:langcross})} is rewritten the frequency domain as: 
\begin{alignat}{2}
\langle \widehat{F_{i}}(\freq_1) \widehat{F}_j^{\ast}(\freq_2)\rangle & \stack{(\ref{eq:fourier})}{=}  2 D_{ij} \delta(\freq_1-\freq_2).
\label{eq:fourierdiffusion} 
\end{alignat}
A usual approach for solving the system from \mbox{Eq.~(\ref{eq:rateposta})} involves small-signal analysis. Small deviations from a \mbox{steady-state value} are assumed: 
\begin{subequations}
	\label{eq:def}
	\begin{alignat}{5}
	\I &\simeq  \meanI +&& \pd &&=  \meanI + &&  \pddev  \qquad \mathrm{with} \quad \meanI \gg \pd  \label{eq:pdef} \\
	\phi &\simeq  && \phip &&= &&\phipdev\label {eq:phidef} \\
	N &\simeq \meanN +&&  \nop &&= \meanN +&&  \nopdev  \qquad \mathrm{with} \quad \meanN \gg \nop,  \label{eq:ndef}
	\end{alignat}
\end{subequations}
where the steady state value of the phase is assumed to be zero.
The full linearization of the rate equations is shown in \mbox{Appendix~\ref{sec:apsmallsignal}}, which uses the following definitions:
\begin{subequations}
	\label{eq:abgammacoefficients}
	\begin{multicols}{2}
		\setlength{\abovedisplayskip}{0pt}
		\setlength{\abovedisplayskip}{0pt}
		\setlength{\belowdisplayskip}{0pt}
		\begin{alignat}{2}
		\kc  &\equiv \ki \tie  \cos\left( \phased\right) \label{eq:acoefficients}\\
		\ks &\equiv \ki \tie \sin\left( \phased\right) \label{eq:bcoefficients} \\
		\mathsf{K}_{s}  & \equiv  \ksr + \ksl \\
		\mathsf{K}_{c}  & \equiv 1 + \kcr - \kcl
		\end{alignat}
		\setlength{\belowdisplayskip}{0pt}
		\setlength{\columnsep}{0em}
		\begin{alignat}{2}
		\gammai &\equiv \R  / \meanI \\
		 \again  & =  \gvg \diffgain \\
		 G_{i} & \approx \again (N_{i} - N_{\mathrm{tr}})\\
		\gamman & \equiv \again  \meanI  + \rrecN, \label{eq:gammaN} 
		\end{alignat}
	\end{multicols}
\end{subequations} 
\noindent where a linear approximation for the gain has been introduced, with $a$ the differential gain coefficient and $N_{\mathrm{tr}}$ the number of electrons at transparency.
Applying the Fourier transform from  \mbox{Eq.~(\ref{eq:fourier})} 
to \mbox{Eqs.~(\ref{eq:prate})}, (\ref{eq:phirate}) \mbox{and (\ref{eq:carrierrate})}, the following system of equations is obtained in the frequency domain: 
\begin{subequations}
	\begin{alignat}{3}
	i \freq \mathsf{K}_{c} \hatpd & \stack{(\ref{eq:prate})}{=}  \, && \, 
	\again   \meanI  \hatnop 
	- \gammai \hatpd 
	- 2 i  \freq \meanI \mathsf{K}_{s} \hatphip 
	 + \hatFI \label{eq:fourier_p} \\
	2 i \freq \mathsf{K}_{c} \hatphip & \stack{(\ref{eq:phirate})}{=}  \, && 
	\ah \again  \hatnop 
	+  i \freq \frac{\mathsf{K}_{s}}{\meanI } \hatpd 
	+ 2 \hatFphi \label{eq:fourier_phi} \\
	i \freq \hatnop& \stack{(\ref{eq:carrierrate})}{=}  \, && - \gamman \hatnop - \go  \hatpd + \hatFN ,\label{eq:fourier_n}
	\end{alignat}
	\label{eq:fouriersystem}
\end{subequations}
where the unknowns $\hatpd$, $\hatphip$, $\hatnop$, and $\hatFI$, $\hatFphi$ and $\hatFN $ depend on the fourier frequency $\freq$. These equations are the first step to obtain the FN PSD.


\subsection{Power spectral density and laser intrinsic linewidth}
\label{sec:linepsd}

The next step is to find an expression for $\hatphip$ from which the FN PSD, and thus the laser intrinsic linewidth, can be computed.
Defining:
\begin{subequations}
	\label{eq:acoefdef}
	\begin{alignat}{2}
	\Aphi & \equiv
	\left(i \freq + \gamman\right) 
	\left(
	i \freq \mathsf{K}_{c} 
	+ \frac{\GnG }{i \freq + \gamman} \meanI  
	+ \gammai\right) \label{eq:aphimoni}\\
	2 \AI  &\equiv i \freq \mathsf{K}_{s} \frac{i \freq + \gamman}{ \meanI } - \ah \GnG \label{eq:ai}\\
	2 \AN  & \equiv \frac{\ah \Aphi +2  \meanI \AI  }{i \freq + \gamman} \again  \label{eq:an} \\
	\deltam & \equiv \mathsf{K}_{c} \Aphi
	-  \ah \GnG  \meanI  \mathsf{K}_{s}                    
	+ i \freq \mathsf{K}_{s}^2 \left(i \freq + \gamman\right), \label{eq:deltamoni_aux}
	\end{alignat}
\end{subequations}
the expression for $\hatphip$, as shown in \mbox{Appendix~\ref{sec:appsd}}, can be written as:
\begin{alignat}{2}
\hatphip & \stack{(\ref{eq:phihatpretty_app})}{=}\frac{
	\AN  \, \hatFN  +  \AI   \, \hatFI  + \, \Aphi \hatFphi}{i \freq \deltam} . \label{eq:phihatpretty}
\end{alignat}
From \mbox{Eq.~(\ref{eq:phihatpretty})} it is possible to calculate an expression for the \mbox{PSD \cite{rowe1965}: }
\begin{equation}
\PSD (\freq) = \frac{\freq^2}{2 \pi^2}\langle |\hatphip(\freq)|^2 \rangle, \label{eq:PSDfromphi} 
\end{equation}
which, using the following definitions:
\begin{subequations}
	\label{eq:lambdasmoni}
	\begin{multicols}{2}
		\setlength{\abovedisplayskip}{0pt}
		\setlength{\belowdisplayskip}{0pt}
		\setlength{\columnsep}{0em}
		\begin{alignat}{2}
		\mathsf{F}_{0} 
		& \equiv \gammai^2  + \mathsf{K}_{c}^2  \gammansq - 2 \mathsf{K}_{c} \GnG   \meanI , \label{eq:upsilon} \\
		\mathsf{F}_{1}  & \equiv \mathsf{K}_{s}^2 + \mathsf{K}_{c}^2  \label{eq:F1}
		\end{alignat}
		\setlength{\abovedisplayskip}{0pt}
		\setlength{\belowdisplayskip}{0pt}
		\setlength{\columnsep}{0em}
		\begin{alignat}{2}
		\mathsf{F}_{2}  & \equiv \mathsf{K}_{c} \ah + \mathsf{K}_{s} \label{eq:F2} \\
		\mathsf{F}_{3}  & \equiv \mathsf{K}_{c} - \ah \mathsf{K}_{s} ,
		\end{alignat}
	\end{multicols}
\noindent and:
	\begin{alignat}{2}
	\Lambda_{4} & \equiv  4 \mathsf{F}_{1}  \dpp \\
	\Lambda_{2} & \equiv \again^2 \mathsf{F}_{2}^2 \dnn
	+ 4 \dpp \left(\mathsf{K}_{s}^2 \gammansq + \mathsf{F}_{0} \right) - 2 \frac{\mathsf{K}_{s} }{\meanI} \dii  \again  \left(\gamman  \mathsf{F}_{2}  - \gammai \ah  -\ah  \go \right) \\
	\Lambda_{0} & \equiv \ah^2 \again^2 \left[\dii \left(\gammai^2 + \go^2 + 2 \gammai \go \right) + \gammai \rrec\right]
	+ \left(\gamman \gammai + \GnG  \meanI \right)^2 
	4 \dpp, \label{eq:lambdazero}
	\end{alignat}
\end{subequations}
can be written as:
\begin{alignat}{2}
2 \pi^2 \PSD & \stack{(\ref{eq:lambdasmoni})}{=} \stack{(\ref{eq:deltaphiomegamoni_app})}{=}  \frac{\Lambda_{4} \freq^4 + \Lambda_{2} \freq^2 + \Lambda_{0}}{2 |\deltam|^2 }. \label{eq:deltaphiomegamoni}
\end{alignat}
Finally, from the following expression:~\cite{Spano_1984} 
\begin{equation}
\PSD (f \to 0) = 2 \pi \Delta f \label{eq:psdlinewidth},
\end{equation}
which is valid for a Lorentzian lineshape,
the intrinsic linewidth can be obtained.
As shown in \mbox{Appendix~\ref{sec:aplinewidth}}:
\begin{alignat}{2}
F \equiv \frac{\Delta f}{\Delta f_{0} \left( 1 + \ah^2 \right)} 
& \stack{(\ref{eq:correctingAgrawal_app})}{=}  \left[1+ \gamma_{_{\mathrm{H}}} \kr \tre \cos \left( \phasedr + \theta_{_{\mathrm{H}}}\right) - \gamma_{_{\mathrm{H}}} \kl \tle  \cos \left( \phasedl - \theta_{_{\mathrm{H}}} \right)\right]^{-2},\label{eq:correctingAgrawal}
\end{alignat}
where $\Delta f_0$ is the Schawlow–Townes linewidth~\cite{Schawlow1958}. The expression found for the intrinsic linewidth has two feedback terms that account for one contribution from each side, with a sign that depends on $\phasedl$ and $\phasedr$. Recalling from \mbox{Eq.~(\ref{eq:phased})} that these quantities are a function of $\tie$ and $\phim$, a proper design of the laser can yield linewidth stability or a reduction of the intrinsic linewidth with respect to the case of one-sided feedback. This is further explored using a numerical analysis in \mbox{Sec.~\ref{sec:sims}}.
Additionally, it is possible to find an analytical expression for \mbox{Eq.~(\ref{eq:correctingAgrawal})} under the conditions for frequency stability, namely \mbox{conditions (\ref{eq:kappa_assumption})}, (\ref{eq:text_assumption}) \mbox{and (\ref{eq:phim_condition})}. Using this assumptions in \mbox{Eq.~(\ref{eq:correctingAgrawal})}:
\begin{alignat}{2}
 F & =  \left\{1+ \gamma_{_{\mathrm{H}}} \kappa \taue \left[\cos \left( \phasedr + \theta_{_{\mathrm{H}}}\right) - \cos \left( \phasedl - \theta_{_{\mathrm{H}}} \right) \right] \right\}^{-2}. \label{eq:linewidth_stable_case}
\end{alignat}
Taking a closer look at the feedback terms yields:
\begin{alignat}{2}
\cos \left( \phasedr + \theta_{_{\mathrm{H}}}\right)
 \stack{(\ref{eq:phim_condition})}{=} \wref \taue  + \phi_{m_{1}} - 2 \theta_{_{\mathrm{H}}} + \theta_{_{\mathrm{H}}}
 = \cos \left( \phasedl - \theta_{_{\mathrm{H}}}\right). \label{eq:linewidthcos}
\end{alignat}
Using  \mbox{Eq.~(\ref{eq:linewidthcos})} in \mbox{Eq.~(\ref{eq:linewidth_stable_case})} yields a value of $F=1$ which indicates that, under the assumed conditions, the intrinsic linewidth is insensitive to feedback. This result is significant as under the same condition the frequency is also feedback-insensitive, as shown in \mbox{Sec.~\ref{sec:ampphase}}, regardless of lasing frequency. It is worth noting however that weak feedback was assumed in this analysis, with which the upper bound for feedback strength under which these equation are valid is not established. Nevertheless, achieving stability in the full frequency domain even under this condition is an improvement with respect to the single feedback case.

\section{Numerical study}
\label{sec:sims}

Laser stability is studied by numerically evaluating the equations for the shift in lasing frequency, threshold gain and intrinsic linewidth under the revised EOF conditions, namely Eqs.~(\ref{eq:gainfreqshift}) \mbox{and~(\ref{eq:correctingAgrawal})}, under different feedback parameters. Particular attention is given to the previously analyzed case under \mbox{conditions (\ref{eq:kappa_assumption})}, \mbox{(\ref{eq:text_assumption})} \mbox{and (\ref{eq:phim_condition})}, which shows stable solutions for the lasing frequency and intrinsic linewidth. System tolerances to each of these conditions are explored by varying each while keeping the other two fixed. The simulated equations are plotted as a function of the unperturbed laser frequency multiplied by $\tre$ which, given the periodicity of the functions, is thus kept between 0 and 1 (\textit{i.e} $\wref \tre \in (0,2\pi)$). Additionally, simulations assume $\ah = 3$. This value is compatible with measurements for semiconductor lasers~\cite{Yu_2004}, and \mbox{thus meeting condition (\ref{eq:phim_condition})} requires that $\Delta \phi_{\mathrm{m}} = 2\theta_{_\mathrm{H}} \simeq 2.5$.

 Results are compared with the case with feedback from a single side, in which parameter:
\begin{equation}
\kl = 0. \label{eq:single_feedback_condition}
\end{equation}
In this case, as shown in~\cite{Tkach1986_1}, as feedback strength increases, solutions for the lasing frequency become multi-valued. This gives rise to instabilities in the system such as mode hopping or coherence collapse regimes. The separation between single-valued and multi-valued solutions is related to the coefficient:
\begin{equation}
C = \gamma_{_{\mathrm{H}}} \kr \tre, \label{eq:C}
\end{equation}
where $C=1$ is the critical value that separates both behaviours.

\begin{figure}[h]
	\begin{center}
		\includegraphics[width=0.75\linewidth]{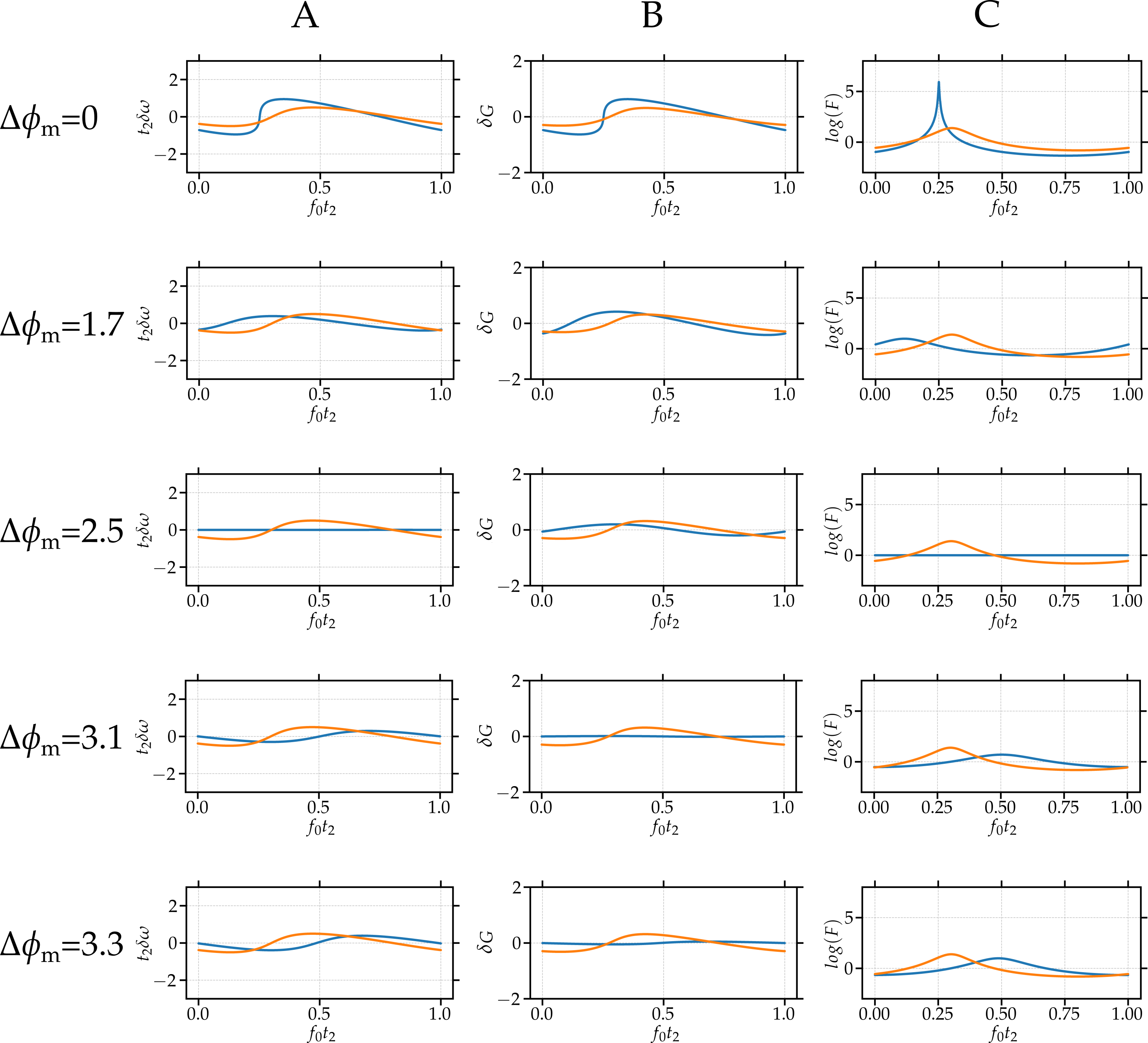}
		\caption{Simulations for $C=0.5$ under conditions (\ref{eq:kappa_assumption}) and (\ref{eq:text_assumption}) for different $\Delta \phi_{\mathrm{m}}$. Full solution shown in blue, single feedback case shown in orange. Column A shows the lasing frequency shift results. Column B shows the threshold gain shift. Column C shows the intrinsic linewidth variations.}
		\label{fig:cless1}
	\end{center}
\end{figure}

Simulation results of the proposed system under \mbox{conditions (\ref{eq:kappa_assumption})} \mbox{and (\ref{eq:text_assumption})} are thus compared to the single feedback case for three cases:

\textbf{Case 1:} $C=0.5$. This represents the single-feedback case with a single solution, and results for various values of $\Delta \phi_{\mathrm{m}}$ are shown in Fig.~\ref{fig:cless1}.  Column A, B and C show the variations in lasing frequency, threshold gain and intrinsic linewidth respectively, and the blue and orange plots represent the full system and the single-feedback case respectively. These labels are maintained throughout the document.
The upper row shows the case where $\Delta \phi_{\mathrm{m}} = 0$, \textit{i.e.} there is no additional phase difference between the external cavities. Linewidth narrowing for a wide range of frequencies can be observed, evidenced by negative values, whose magnitude is higher than in the single feedback case. Additionally, the signal is singled valued in the full domain, yet it is close to the critical point where multi-valued solutions arise.
As $\Delta \phi_{\mathrm{m}}$ increases, the amplitude of the lasing frequency shift is reduced until becoming zero for all values when condition (\ref{eq:phim_condition}) is met, as expected from previous analysis. Under this condition the intrinsic linewidth does not experience fluctuations either, and the amplitude of the threshold gain fluctuations is lower than in the single feedback case, indicating better stability across the three analyzed parameters with respect to the single feedback case. For the case of $\Delta \phi_{\mathrm{m}}  = \pi$ the threshold gain shows no fluctuations as predicted by \mbox{Eq.~(\ref{eq:gainpi})}, and while the lasing frequency and intrinsic linewidth fluctuations are no longer zero, they are less pronounced than in the single feedback case. Further increases in $\Delta \phi_{\mathrm{m}}$ show an increase in the fluctuations across all functions, and for $\phi_{\mathrm{m}} > 6$ multi-valued solutions arise.

\begin{figure}[h]
	\begin{center}
		\includegraphics[width=0.75\linewidth]{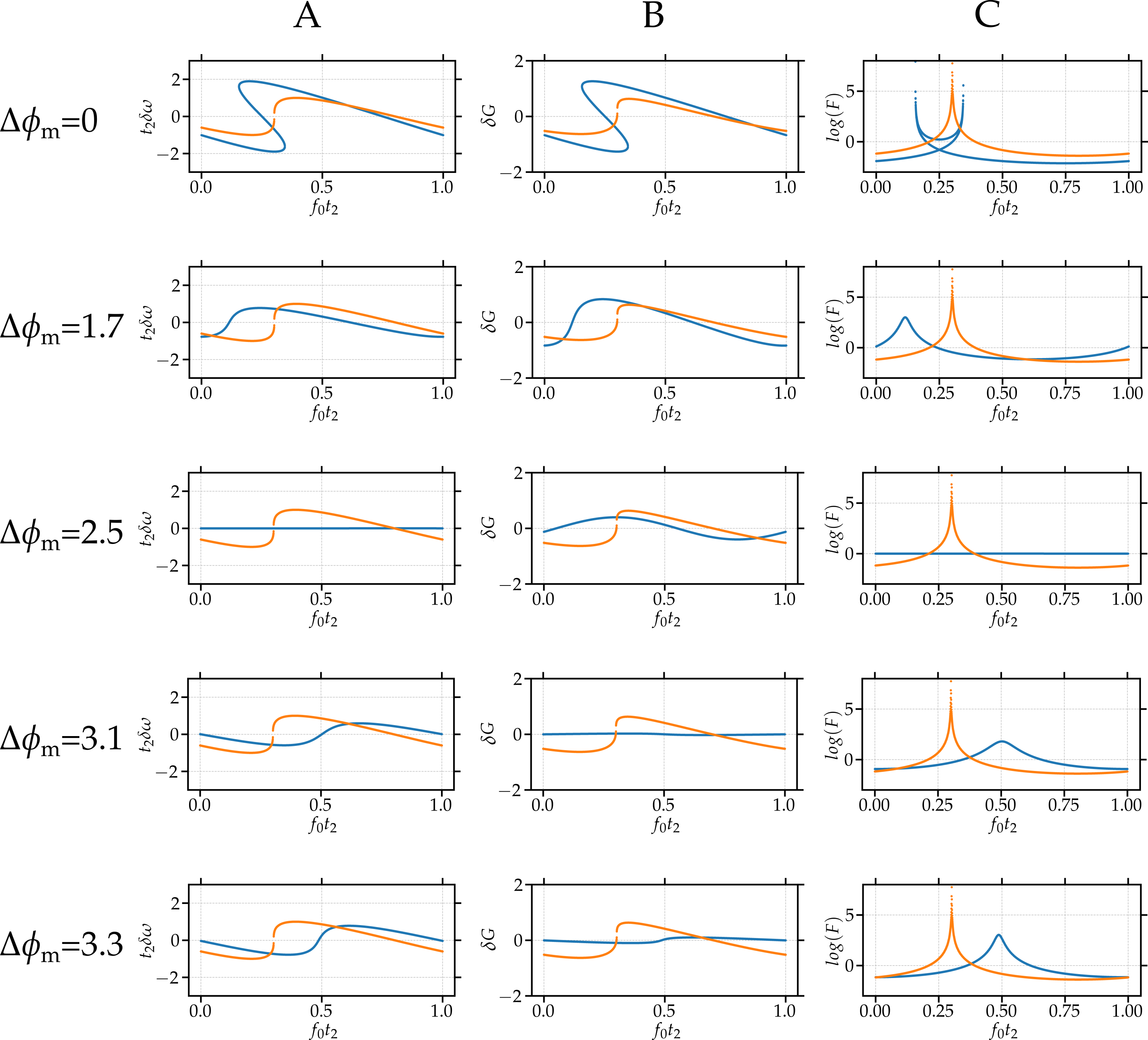}
		\caption{Simulations for $C=1$ under conditions (\ref{eq:kappa_assumption}) and (\ref{eq:text_assumption}) for different $\Delta \phi_{\mathrm{m}}$. Full solution shown in blue, single feedback case shown in orange. Column A shows the lasing frequency shift results. Column B shows the threshold gain shift. Column C shows the intrinsic linewidth variations.}
		\label{fig:cequal1}
	\end{center}
\end{figure}

\textbf{Case 2:} $C=1$. This represents the limiting case between single and multivalued solutions in the single-feedback case. Results for various values of $\Delta \phi_{\mathrm{m}}$ are shown in Fig.~\ref{fig:cequal1}.
As expected, the threshold gain is stable for $\Delta \phi_{\mathrm{m}} = \pi$, and meeting  \mbox{condition (\ref{eq:phim_condition})} results in a stable lasing frequency and intrinsic linewidth. As $\Delta \phi_{\mathrm{m}}$ deviates from these optimal points in either direction, the amplitude of fluctuations increase until reaching multi-valued solutions for $\Delta \phi_{\mathrm{m}} < 1.5$ and $\Delta \phi_{\mathrm{m}} > 3.5$. 
Comparing these results with the previous case shows that as feedback increases, the single valued solutions become more sensitive to the value of $\Delta \phi_{\mathrm{m}}$.

\begin{figure}[h]
	\begin{center}
		\includegraphics[width=0.75\linewidth]{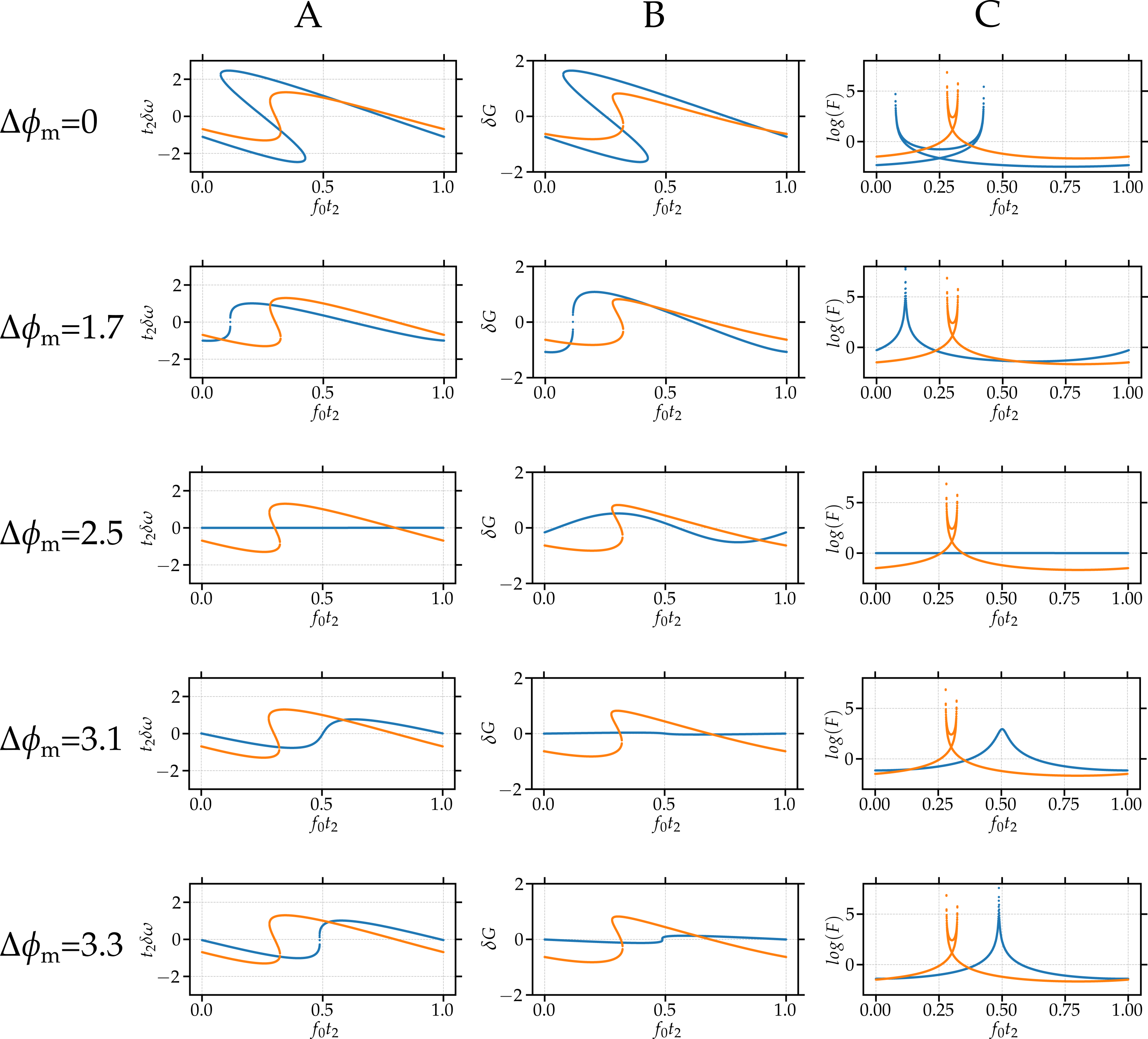}
		\caption{Simulations for $C=1.3$ under conditions (\ref{eq:kappa_assumption}) and (\ref{eq:text_assumption}) for different $\Delta \phi_{\mathrm{m}}$. Full solution shown in blue, single feedback case shown in orange. Column A shows the lasing frequency shift results. Column B shows the threshold gain shift. Column C shows the intrinsic linewidth variations.}
		\label{fig:cmore1}
	\end{center}
\end{figure}

\textbf{Case 3:} $C=1.3$. This represents the single-feedback case with multi-valued solutions. Results for various values of $\Delta \phi_{\mathrm{m}}$ are shown in Fig.~\ref{fig:cmore1}.
\noindent In the single feedback case, multi-valued solutions are present in a given frequency range, and this span increases with increasing feedback strength. The multi-valued characteristics are evidenced experimentally with unstable regimes characterized by mode hopping and eventually coherence collapse for sufficiently high feedback. In contrast, the system proposed in this work shows that by tuning the value of $\Delta \phi_{\mathrm{m}}$ to meet condition (\ref{eq:phim_condition}), even with increasing feedback it is possible to achieve stable performance regardless of frequency.
In the case shown in Fig.~\ref{fig:cmore1} for $C=1.3$, single valued solutions can be found for $\Delta \phi_{\mathrm{m}} \in (1.5,3.3)$ which is equivalent to a phase variation of more than $90^{\circ}$. Still, comparing with previous cases it is possible to see that as feedback increases, the single valued solutions tolerance with respect to $\Delta \phi_{\mathrm{m}}$ decreases. Nevertheless, it is an improvement with respect to the single feedback case which shows no single value solutions across all frequencies for $C>1$.

Taking all cases into account, it can be seen that linewidth narrowing regions are present for all cases of $\Delta \phi_{\mathrm{m}}$ analyzed. The only exception are the stable cases when meeting the three conditions (\ref{eq:kappa_assumption}), (\ref{eq:text_assumption}) and (\ref{eq:phim_condition}), yet the feedback-insensitivity provided by this case is also beneficial. Selecting an appropriate $\Delta \phi_{\mathrm{m}}$ can thus be used to harness linewidth narrowing properties at a desired frequency. 

Furthermore, system tolerances to conditions (\ref{eq:kappa_assumption}) and (\ref{eq:text_assumption}) are studied, while maintaining condition (\ref{eq:phim_condition}). Results for 20\% deviation are shown in Figs.~(\ref{fig:tolcless1}),~(\ref{fig:tolcequal1}) and (\ref{fig:tolcmore1}) for cases 1, 2 and 3 respectively, where single valued solutions are found in all cases.
Results show a high tolerance with respect to feedback strength. Looking at case 3, single valued solutions are obtained for $\kl/\kr \in(0.2, 1.7)$. While the system is no longer feedback-insensitive, results evidence single-valued solutions that are robust with respect to condition  (\ref{eq:kappa_assumption}).                                                                                                                                     This stable case is more sensitive with respect to (\ref{eq:text_assumption}), with single valued solutions achieved within $\tle/\tre \in(0.47, 1.2)$.  Nevertheless, tolerances become once again stricter as feedback increases for both parameters, thus laser design is of paramount importance and should focus on meeting the discussed stability conditions.
In particular, choosing equal lengths for both external cavities should be enough to meet condition (\ref{eq:text_assumption}). Fabrication tolerances in foundry processes are the main limiting factor for time delay accuracy.
To meet condition (\ref{eq:kappa_assumption}), a possible approach is to merge the output of the two external cavities into a single one using a coupler, which can be included in the laser on-chip.
Finally, as mentioned before, condition (\ref{eq:phim_condition}) can be met by using a phase shifter, which is a mature component in active platforms.

All in all, results demonstrate that, with proper design of the laser cavity the conditions (\ref{eq:kappa_assumption}), (\ref{eq:text_assumption}) and (\ref{eq:phim_condition}) can be met, with which it is possible to obtain lasing frequency and intrinsic linewidth insensitivity to feedback.

\section{Discussion}

The existing mature photonic integration fabrication processes are very flexible with respect to device geometry. They are however limited by a lack of commercially available on-chip isolators, and thus new approaches are required to minimize the effect feedback has on laser stability.
The current work proposes a theoretical extension of laser dynamics under EOF by considering two external reflections, one from each side of the cavity, instead of the single feedback approach explored in previous literature.
The proposed analysis yields new laser dynamic equations. These are numerically solved, which show the existence of a stable regime in terms of lasing frequency and intrinsic linewidth, with high feedback tolerances.
Moreover, feedback-insensitivity is achievable under conditions that can be met with current laser fabrication processes and components, such as phase shifters for meeting condition (\ref{eq:phim_condition}) and couplers for condition ~(\ref{eq:kappa_assumption}). 
It has previously been shown, for the single feedback case, that tuning the feedback phase can result in linewidth narrowing ~\cite{Zhao2018}, however this still requires low feedback levels and a precise phase shift which can suffer variations due to external parameters, such as temperature or driving currents. In the proposed approach, the stability conditions do not require specific values, instead relating the feedback parameters from one side to those from the other side. For example, condition (\ref{eq:text_assumption}) only implies equal roundtrip times at the external cavities, regardless of value. This allows for additional flexibility in the feedback parameters and gives versatility to the device. Furthermore, this method allows for feedback-insensitivity across the full spectra, which is not seen in the single feedback case. This is of particular importance for tunable lasers, as all lasing frequencies are thus equally affected. Additionally, it relaxes the need for an isolator, reducing the cost and size of packaging processes.
Another significant improvement of the proposed method with respect to the single feedback case is the increase in feedback tolerance: higher levels of feedback strength are allowed without seeing multi-valued solutions, which result in mode-hopping seen experimentally. As a weak feedback approximation is used, the upper bound for feedback tolerance cannot be extracted from this analysis. Despite this, even under this approximation, tolerances are higher than that of the single feedback case. Furthermore, this system has a high tolerance to deviations from the optimal stability conditions as analyzed in the previous section. Linewidth narrowing can be achieved in these cases, for certain frequency values which can be tuned by selecting appropriate feedback parameters, as was the case for single feedback conditions, while maintaining stable solutions.

Finally, while the dynamics under consideration are complex, the laser system itself involves a straight-forward configuration using widely used components, which are available in generic foundry platforms. Previously studied methods to reduce feedback sensitivity include resourceful yet intricate designs. The proposed system is, in contrast, potentially easier to design, fabricate and characterize. An experimental study of this laser system is essential to validate the obtained results, and more importantly to explore the limitations of the model, and is the next step for a more comprehensive understanding of the proposed system.


\section{Conclusions}
This work explores an extension of the theoretical background of EOF. By assuming that feedback couples into the laser cavity from both sides, new dynamic equations are found for the lasing frequency, the threshold gain and the intrinsic linewidth. These are numerically evaluated to analyze laser stability. Results show the existence of a stable solution, with feedback-insensitive lasing frequency and intrinsic linewidth, regardless of the lasing frequency. This case is obtained by tuning the phase of the feedback field, for external cavities with equal lengths and coupling factors. Furthermore, the feedback-insensitive case exists regardless of the feedback strength,within a weak feedback approximation, which is an major improvement with respect to the single feedback case. Furthermore, the stability conditions shows good tolerances with respect to all feedback parameters, albeit they become stricter as the feedback strenght increases. Additionally, solutions with linewidth reduction are observed. In particular, choosing feedback parameters close to the feedback-insensitive conditions ensure stable solutions that are feedback tolerant. Finally, the proposed system relies on few components in straight forward configurations, and the stable conditions can be met with mature components available in generic foundry platforms. This enables close to market, low cost, feedback tolerant semiconductor lasers and has has direct applications in multiple fields which rely on stable laser sources, such as coherent communications and spectroscopy.


\vspace{6pt}

\appendix
\numberwithin{equation}{section}

\section{EFFECTIVE REFLECTION COEFFICIENTS}

\label{ap:rhoef}

Consider the slowly-varying amplitudes $\mathcal{A}$ shown in \mbox{Fig.~\ref{fig:lasercav}}, where a slowly varying electric field is assumed:
\begin{align}
\mathcal{E}(t) = \mathcal{A}(t) e^{-i \wfb t},\label{eq:fielddef}
\end{align}
in which the lasing frequency with feedback is $f_{\mathrm{FB}} = \frac{\wfb}{2 \pi}$. The relation between $\mathcal{A}_{1}^{'\pm}$ and $\mathcal{A}_{1}^{\pm}$ is: 
\begin{subequations}
	\label{eq:field12}
	\setlength{\abovedisplayskip}{0pt}
	\setlength{\belowdisplayskip}{0pt}
	\setlength{\columnsep}{0em}
	\begin{multicols}{2} 
		\begin{align}
		´		\mathcal{A}_{1}^{'-} e^{-i \phasedr/2} & = \mathcal{A}_{1}^{-} \label{eq:field1}
		\end{align}
		\setlength{\abovedisplayskip}{0pt}
		\setlength{\belowdisplayskip}{0pt}
		\begin{alignat}{2}
		\mathcal{A}_{1}^{'+} & = \mathcal{A}_{1}^{+} e^{-i \phasedr/2} \label{eq:field2}.  
		\end{alignat} 
	\end{multicols}
\end{subequations}
\noindent Using the 
transmission coefficient:
\begin{align}
\tau_{j}^2 & = 1 - |\rhoi|^2, \label{eq:transmission}
\end{align}
\noindent  where $j = 1,2$, the equations that describe the fields propagation are obtained by considering the reflections and transmissions at the interfaces:
\begin{subequations} 
	\label{eq:fieldFP}
	\setlength{\abovedisplayskip}{0pt}
	\setlength{\belowdisplayskip}{0pt}
	\setlength{\columnsep}{0em}
	\begin{multicols}{2}
		\begin{align}
		\mathcal{A}_{0}^{-} & = - \rhor  \mathcal{A}_{0}^{+} + \tau_{2}  \mathcal{A}_{1}^{-} \label{eq:kfp1}\\
		\mathcal{A}_{1}^{+} & = - \rhor  \mathcal{A}_{1}^{-} + \tau_{2}  \mathcal{A}_{0}^{+} \label{eq:kfp2}
		\end{align}
		\setlength{\abovedisplayskip}{0pt}
		\setlength{\belowdisplayskip}{0pt}
		\begin{alignat}{2}
		\mathcal{A}_{1}^{'-} & = - \rer  \mathcal{A}_{1}^{'+} \nonumber \\
		\stack{(\ref{eq:field12})}{\Longrightarrow}  \mathcal{A}_{1}^{-} e^{i \phasedr/2} & = \rer  \mathcal{A}_{1}^{+} e^{-i \phasedr/2} \label{eq:kfp3}.
		\end{alignat}
	\end{multicols}
\end{subequations}
\noindent Inserting \mbox{Eq.~(\ref{eq:kfp3})} into \mbox{Eq.~(\ref{eq:kfp1})} \mbox{and~(\ref{eq:kfp2})}: 
\begin{subequations} 
	\label{eq:fieldFP_aux1}
	\setlength{\abovedisplayskip}{0pt}
	\setlength{\belowdisplayskip}{0pt}
	\setlength{\columnsep}{0em}
	\begin{multicols}{2}
		\begin{align}
		\mathcal{A}_{0}^{-} & = - \rhor  \mathcal{A}_{0}^{+} - \tau_{2}  \rer  e^{-i \phasedr} \mathcal{A}_{1}^{+} \label{eq:kfp21}\\
		\mathcal{A}_{1}^{+} & =  \rhor \rer  e^{-i \phasedr} \mathcal{A}_{1}^{+} + \tau_{2}  \mathcal{A}_{0}^{+}
		\end{align}
		\setlength{\abovedisplayskip}{-3pt}
		\setlength{\belowdisplayskip}{0pt}
		\begin{alignat}{2}
		& \nonumber \\
		\Longrightarrow \qquad \mathcal{A}_{1}^{+}& =  \frac{\tau_{2}}{1-\rhor \rer e^{-i \phasedr} }  \mathcal{A}_{0}^{+}, \label{eq:kfp22}
		\end{alignat}
	\end{multicols}
\end{subequations}
\noindent and replacing \mbox{Eq.~(\ref{eq:kfp22})} in \mbox{Eq.~(\ref{eq:kfp21})}:
\begin{alignat}{2}
\mathcal{A}_{0}^{-} & = - \rhor  \mathcal{A}_{0}^{+} -   \frac{\tau_{2}^2 \rer }{1-\rhor \rer e^{-i \phasedr/2} }  \; e^{-i \phasedr/2} \mathcal{A}_{0}^{+} \nonumber \\
\stack{(\ref{eq:weak_feedback})}{\Longleftrightarrow} \mathcal{A}_{0}^{-} / \mathcal{A}_{0}^{+} \equiv \refr & \stack{(\ref{eq:transmission})}{=} \stack{(\ref{eq:kappa})}{=} \rhor \left(1 + \kr \tcav e^{-i \phasedr}\right).
\label{eq:effectiveright}
\end{alignat}
An analogous analysis can be done for \mbox{mirror 1}.
The relations from \mbox{Eq.~(\ref{eq:fieldFP})} remain the same, as the equations for field transmission and reflection are valid in either case.
However, the phase induced by the external cavity, represented by \mbox{Eq.~(\ref{eq:field12})}, has an opposite sign given that the fields travel in opposite directions:
\begin{subequations}
	\label{eq:field12left}
	\setlength{\abovedisplayskip}{-10pt}
	\setlength{\belowdisplayskip}{-10pt}
	\setlength{\columnsep}{0em}
	\begin{multicols}{2}
		\begin{align}
		\mathcal{A}_{1}^{'-} e^{i \phasedl /2} & = \mathcal{A}_{1}^{-} \label{eq:field1left}
		\end{align}
		\setlength{\abovedisplayskip}{-10pt}
		\setlength{\belowdisplayskip}{-10pt}
		\begin{alignat}{2}
		\mathcal{A}_{1}^{'+} & = \mathcal{A}_{1}^{+} e^{i \phasedl/2} \label{eq:field2left}, 
		\end{alignat}
	\end{multicols}
\end{subequations} 
\noindent and thus: 
\begin{alignat}{2}
\refl & = \rhol \left(1 + \kl \tcav e^{i \phasedl}\right).
\label{eq:effectiveleft}
\end{alignat}
%
%
%
%
\mbox{Eqs.~(\ref{eq:effectiveright})} \mbox{and~(\ref{eq:effectiveleft})} can be summarized as:
%
%
%
\begin{alignat}{2}
\refi  \Big/ \rhoi &  \stack{(\ref{eq:effectiveright})}{=} \stack{(\ref{eq:effectiveleft})}{=}  1 + \ki \tcav e^{\pm i\phased}
= 1 + \ki \tcav \cos(\phased) 
\pm i \ki \tcav \sin(\phased). \label{eq:rhoFinal_app}
\end{alignat}
where the plus and minus signs corresponds to $j = 1$ (left mirror) and $j = 2$ (right mirror), respectively.  Both effective reflection coefficients are considered in the analysis to obtain the laser dynamic equations.


\section{AMPLITUDE AND PHASE CONDITIONS}
\label{ap:conditions}

To obtain the revised lasing conditions resulting from the additional feedback term, extracting the amplitude and phase of the effective reflection coefficients is needed. 
In polar notation:
\begin{align}
\refi  =  \left|\refi\right|e^{i \rphii},	\label{eq:rhopolar}
\end{align}
where the magnitude is computed as:
\begin{alignat}{2}
\left|\refi\right|^2   \bigg/    \rhoi^2 & \stack{(\ref{eq:rhoFinal_app})}{=}  
\left[1 + \ki \tcav\cos(\phased) \right]^2 
+  \left[\ki \tcav\sin(\phased)\right] ^2  \nonumber \\
 &\, \,  =  1 + 2\ki \tcav\cos(\phased) + \ki^2\tcav^2. \label{eq:aux_amp}
\end{alignat}
Assuming that the external feedback is weak compared to the internal one:
\begin{equation}
\left|\rei\right| \ll \left|\rhoi\right| \quad  \stack{(\ref{eq:kappa})}{\Rightarrow} \quad \ki \tcav \ll 1,
\label{eq:approximation}
\end{equation}
the last  term on the right-hand side of \mbox{Eq.~(\ref{eq:aux_amp})} can be neglected, resulting in:
\begin{alignat}{2}
%
\left. \left|\refi (\wfb)\right| \, \right/ \rho_i 
& =   \sqrt{1 + 2 \ki \tcav\cos(\phased)} \stack{(\ref{eq:approximation})}{\approx} 1 + \ki \tcav\cos(\phased)\label{eq:rho_real}.
\end{alignat}
The phase of the effective reflection coefficient is extracted from:
\begin{alignat}{2}
\rphii &= \arctan \left[\left(\refi\right)'' \, \big/ \, \left(\refi\right)'\right] \stack{(\ref{eq:rhoFinal_app})}{=}
\arctan \left[\frac{\pm \ki \tcav\sin(\phased)}{1 + \ki \tcav\cos(\phased)}\right] \nonumber\\
&\stack{(\ref{eq:approximation})}{\approx} \arctan [\pm \ki \tcav\sin(\phased)] \stack{(\ref{eq:approximation})}{\approx} \pm \ki \tcav \sin(\phased).\label{eq:rho_imag}
\end{alignat}
The fields traveling forward and backward in the laser cavity, $\mathcal{E}_{\mathrm{f}}$ and $\mathcal{E}_{\mathrm{b}}$ shown in \mbox{Fig.~\ref{fig:effective_cav}}, can now be related by the effective reflection coefficients: 
\begin{subequations}
	\label{eq:2}
	\setlength{\abovedisplayskip}{0pt}
	\setlength{\belowdisplayskip}{0pt}
	\setlength{\columnsep}{0em}
	\begin{multicols}{2}
		\begin{align}
		\mathcal{E}_{\mathrm{f}}(z=0) &= \refl \, \mathcal{E}_{\mathrm{b}}(z=0)\label{eq:condition0}
		\end{align}
		\setlength{\abovedisplayskip}{0pt}
		\setlength{\belowdisplayskip}{0pt}
		\begin{alignat}{2}
		\mathcal{E}_{\mathrm{b}}(z=L) &= \refr \, \mathcal{E}_{\mathrm{f}}(z=L).\label{eq:conditionL}
		\end{alignat}
	\end{multicols}
\end{subequations}
\noindent Using the propagation constant from \mbox{Eq.~(\ref{eq:beta})},
the fields can be written as:
\begin{subequations}
	\label{eq:3}
	\setlength{\abovedisplayskip}{0pt}
	\setlength{\belowdisplayskip}{0pt}
	\setlength{\columnsep}{0em}
	\begin{multicols}{2}
		\begin{align}
		\mathcal{E}_{\mathrm{f}} &= A_f e^{-i\beta z + \frac{1}{2}(\Gamma g-\alpha)z}\label{eq:forwardfield}
		\end{align}
		\setlength{\abovedisplayskip}{0pt}
		\setlength{\belowdisplayskip}{0pt}
		\begin{alignat}{2}
		\mathcal{E}_{\mathrm{b}} &= A_b e^{-i\beta (L-z) + \frac{1}{2}(\Gamma g-\alpha)(L-z)}\label{backwardfield},
		\end{alignat}
	\end{multicols}
\end{subequations}
\noindent where
$g$ is the gain coefficient.
Replacing \mbox{Eq.~(\ref{eq:2})} into \mbox{Eq.~(\ref{eq:3})}:
\begin{subequations}
	\setlength{\abovedisplayskip}{0pt}
	\setlength{\belowdisplayskip}{0pt}
	\setlength{\columnsep}{0em}
	\begin{multicols}{2}
		\begin{align}
		\mathcal{E}_{f0} & \stack{(\ref{eq:rhopolar})}{=} \left|\refr\right|e^{i \rphir} A_b e^{-i\beta L + (\Gamma g-\alpha) L / 2} \label{eq:fieldincondition0}
		\end{align}
		\setlength{\abovedisplayskip}{0pt}
		\setlength{\belowdisplayskip}{0pt}
		\begin{alignat}{2}
		\mathcal{E}_{b0} & \stack{(\ref{eq:rhopolar})}{=} \left|\refl\right|e^{i \rphil} A_f e^{-i\beta L + (\Gamma g-\alpha) L / 2} , \label{eq:fieldinconditionL}
		\end{alignat}
	\end{multicols}
\end{subequations}
\noindent and inserting \mbox{Eq.~(\ref{eq:fieldincondition0})} into \mbox{Eq.~(\ref{eq:fieldinconditionL})} results in: 
\begin{alignat}{2}
1 & = \left|\refl\right|e^{i \rphil} \left|\refr\right|e^{i \rphir} e^{-2i\beta L + (\Gamma g-\alpha) L} 
\; = \; \left|\refl\refr\right| e^{-i (2 \beta L - \rphil - \rphir)}  e^{(\Gamma g-\alpha) L}. 	\label{eq:preliminarphaseamp}
\end{alignat}
Once lasing has been established, the gain assumes its threshold value:
\begin{figure}[t]
	\begin{center}
		\includegraphics[width=0.7\linewidth]{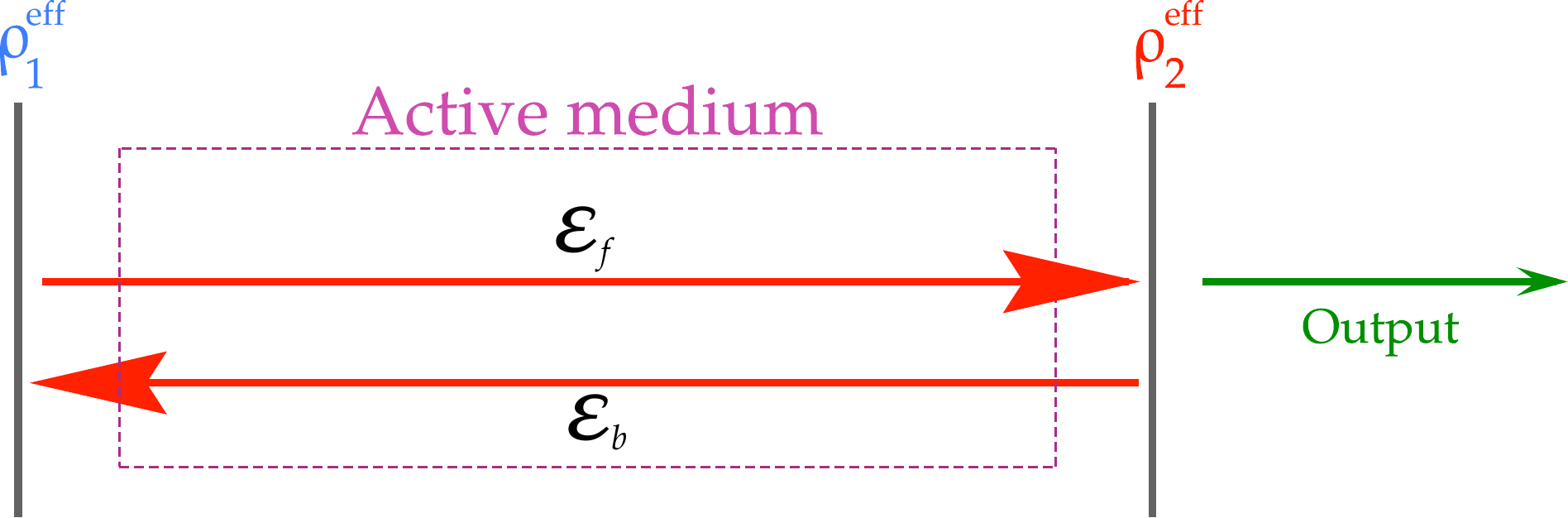}
		\caption{Schematic of the effective cavity of the laser, resulting from calculating effective reflection coeffitients. }
		\label{fig:effective_cav}
	\end{center}
\end{figure}
\begin{alignat}{2}
g &= \gain, \label{eq:gainthreshfeed}
\end{alignat}
where $\gain$ is the threshold gain with feedback. 
Thus, \mbox{Eq.~(\ref{eq:preliminarphaseamp})} yields a lasing condition for the amplitude:
\begin{alignat}{2}
1 & = \left|\refl\refr\right|e^{\left(\Gamma \gain -\alpha \right) L} \nonumber \\
&\stack{(\ref{eq:gainthreshfeed})}{=} \stack{(\ref{eq:rho_real})}{\approx} \rhor \rhol \left[1 + \kr \tcav \cos(\phasedr)\right] \left[1 + \kl \tcav \cos(\phasedl)\right] e^{\left(\Gamma \gain-\alpha \right) L}  \label{eq:ampexplicit_app},
\end{alignat}
and the phase:
\begin{alignat}{2}
2\pi m   &=   2\beta L - \rphir - \rphil   
\stack{(\ref{eq:rho_imag})}{=}  2\beta L + \kr\tcav \sin(\phasedr) - \kl \tcav \sin(\phasedl)\quad , \quad m\in \mathbb{Z}.\label{eq:phaseexplicit_app}
\end{alignat}
where the influence of feedback gives rise to two terms in \mbox{Eqs.~(\ref{eq:ampexplicit_app})} \mbox{and~(\ref{eq:phaseexplicit_app})}, one from each side. The new lasing conditions result in a variation of the lasing frequency and threshold gain of the system, and thus have to be studied to determine the laser dynamics.


\section{THRESHOLD GAIN REDUCTION AND LASING FREQUENCY SHIFT}

\label{ap:threshfreqshift}
Under feedback conditions, new lasing conditions are found which subsequently result in a shift in the laser threshold gain and lasing frequency with respect to the no feedback case.
Without feedback:
\begin{alignat}{2}
\ki = 0, \label{eq:nofeedback}
\end{alignat}
the amplitude condition from \mbox{Eq.~(\ref{eq:ampexplicit_app})} becomes:
\begin{alignat}{2}
1  \stack{(\ref{eq:nofeedback})}{=} \rhor \rhol e^{\left( \Gamma g_{\mathrm{th}} - \alpha\right) L}. \label{eq:ampcondnofeed}\vspace{-3mm}
\end{alignat}
Using the following expansion:
\begin{align}
\ln \left(1 + x\right) &\simeq x, \label{eq:taylorln}
\end{align}
\noindent  the threshold gain reduction due to feedback can be found by computing the ratio between \mbox{Eqs.~(\ref{eq:ampexplicit})} \mbox{and~(\ref{eq:ampcondnofeed})}:
\begin{alignat}{2}
1 & = \frac{\rhor \rhol \left[1 + \kr \tcav\cos(\phasedr)\right] \left[1 + \kl \tcav \cos(\phasedl)\right] e^{\left(\Gamma \gain-\alpha \right) L}}{\rhor \rhol e^{\Gamma g_{\mathrm{th}} - \alpha L}}  \nonumber \\
& = \left[1 + \kr \tcav \cos(\phasedr)\right] \left[1 + \kl \tcav \cos(\phasedl)\right] e^{\Gamma \left(\gain-g_{\mathrm{th}} \right) L} \nonumber \\
\stack{(\ref{eq:defforgain})}{\Leftrightarrow} G - G_{\mathrm{th}} & \approx - 2 \kr \cos(\phasedr) - 2 \kl \cos(\phasedl) . \label{eq:gainfromamp_app}
\end{alignat}
The relation between the right hand terms determines the threshold gain reduction, as discussed in \mbox{Sec.~\ref{sec:ampphase}}.

The phase lasing condition from \mbox{Eq.~(\ref{eq:phaseexplicit_app})} yields the lasing frequency shift equation.
Consider the following definitions related to the effective refractive \mbox{index~\cite{Petterman_Book}}:
\begin{subequations}
	\setlength{\abovedisplayskip}{0pt}
	\setlength{\belowdisplayskip}{0pt}
	\setlength{\columnsep}{0em}
	\begin{multicols}{2}
		\begin{align}
		n & \equiv n' + i n'' \label{eq:complexrefrac} \\
		n_{\mathrm{g}} & \equiv n + \omega \frac{\partial n}{\partial \omega} \label{eq:muef}\\
		n'' &\equiv 	-\frac{c G}{2 \omega \vg} , \label{eq:refracgain}
		\end{align}
		\setlength{\abovedisplayskip}{0pt}
		\setlength{\belowdisplayskip}{0pt}
		\begin{alignat}{2}
		\ah &\equiv \Delta n' / \Delta n'' \label{eq:ah}\\
		\frac{\partial n}{\partial N}  & = \frac{\partial n}{\partial n''} \frac{\partial n''}{\partial N} \nonumber\\
		& \stack{(\ref{eq:complexrefrac})}{=} \stack{(\ref{eq:refracgain})}{=} \stack{(\ref{eq:ah})}{=} - \frac{\partial G}{\partial N} \frac{\ah c }{2 \omega \vg´}  \label{eq:mupartialn} 
		\end{alignat}
	\end{multicols} 
\end{subequations}
\noindent where $n_{\mathrm{g}}$ is the group refractive index and $\ah$ the linewidth enhancement \mbox{factor~\cite{Henry1982}}. To find the lasing frequency shift equation, calculating the change in $\beta$ is first needed:
\begin{align}
c \Delta \beta &\stack{(\ref{eq:beta})}{=} \Delta(n \omega)
= \omega \Delta n + n \Delta \omega 
\stack{(\ref{eq:deltaomega})}{=}  \omega \left[
\frac{\partial n}{\partial N} \left( N-N_{\mathrm{th}} \right) + \frac{\partial n}{\partial \omega}\delta \omega 
\right]  + n \delta \omega \nonumber \\
&  \stack{(\ref{eq:muef})}{=} \stack{(\ref{eq:mupartialn})}{=}- \frac{G - G_{\mathrm{th}}}{2 \vg} \ah c   + n_{\mathrm{g}} \delta \omega 
\stack{(\ref{eq:gainfromamp})}{=}
\left[\kr \cos(\phasedr) + \kl \cos(\phasedl)\right] \frac{ \ah c }{\vg} + n_{\mathrm{g}} \delta \omega,
\label{eq:deltan}
\end{align} 
with which:
\begin{align}
2 L \Delta \beta & \stack{(\ref{eq:gainfromamp})}{=}
\left[\kr \cos(\phasedr) + \kl \cos(\phasedl)\right] \ah \tcav  + \tcav \delta \omega,
\label{eq:deltan}
\end{align} 
\noindent Furthermore, using:
\begin{vwcol}[widths={0.4,0.6},sep=0,rule=0]
		\begin{subequations} 
			\label{eq:arctanid}
		\begin{alignat}{2}
		\sin\left[\arctan(x)\right] & = \frac{x}{\sqrt{1+x^2}} \label{eq:sinarctan} \\
		\vspace{1mm}
		\cos\left[\arctan(x)\right] &= \frac{1}{\sqrt{1+x^2}} \label{eq:cosarctan}
		\end{alignat}

		\begin{alignat}{2} 
		\vspace{1mm}
		\sin(x \pm y )& =\sin (x) \cos (y) \pm \cos (x) \sin (y) \label{eq:sumofsines} \\[1em]
		\cos (\theta_{_{\mathrm{H}}}) &
		\stack{(\ref{eq:alphah})}{=} 
		\stack{(\ref{eq:cosarctan})}{=} 
		\stack{(\ref{eq:gammaagr})}{=}     1 / \gamma_{_\mathrm{H}}, \label{eq:cosgammah}
		\end{alignat}
\end{subequations}
\end{vwcol}
\noindent 
the following can be computed: 
\begin{alignat}{2}
\ah \cos (\phased) 
& \stack{(\ref{eq:sinarctan})}{=} \stack{(\ref{eq:gammaagr})}{=} \stack{(\ref{eq:alphah})}{=} \gamma_{_\mathrm{H}}  \sin\left( \theta_{_{\mathrm{H}}}\right) \cos(\phased)  \nonumber \\
&
\stack{(\ref{eq:sumofsines})}{=}   \gamma_{_\mathrm{H}} \left[\sin(\theta_{_{\mathrm{H}}} \pm \phased)  \mp \cos(\theta_{_{\mathrm{H}}}) \sin(\phased)\right]   \nonumber \\
& \stack{(\ref{eq:cosgammah})}{=} \gamma_{_\mathrm{H}} \, \left[\sin\left(\theta_{_{\mathrm{H}}} \pm \phased \right) \mp \sin(\phased) \Big/ \gamma_{_\mathrm{H}}\right]  \nonumber \\
\Leftrightarrow
\ah \cos(\phased) \pm \sin(\phased) 
& = \gamma_{_\mathrm{H}} \sin \left(\theta_{_{\mathrm{H}}} \pm \phased  \right). \label{eq:sinalphacos}
\end{alignat}
Finally, using the phase condition in \mbox{Eq.~(\ref{eq:phaseexplicit_app})} and assuming without generality loss that $m=0$:
\begin{alignat}{2}
2\pi m 
%
%
%
%
& \stack{(\ref{eq:deltan})}{=}
\tcav \delta \omega 
+ \kr \tcav\left[ \ah \cos(\phasedr)  + \sin(\phasedr)  \right]
+ \kl \tcav \left[ \ah \cos(\phasedl)  -  \sin(\phasedl) \right]\nonumber  \\
\stack{m=0}{\Leftrightarrow}  \delta \omega  & =
- \kr \left[ \ah \cos(\phasedr)  +  \sin(\phasedr) \right]
	+ \kl \left[ \ah \cos(\phasedl) - \sin(\phasedl)  \right] \nonumber
\\
 \delta \omega  &
\stack{(\ref{eq:sinalphacos})}{=}
- \gamma_{_\mathrm{H}} \left[\kr \sin (\phasedr + \theta_{_{\mathrm{H}}}) + \kl \sin (\phasedl - \theta_{_{\mathrm{H}}})\right] \label{eq:omegafeedback_app}
\end{alignat}
which describes the lasing frequency shift as a function of the feedback parameters  $\kl, \kr, \phasedl$ and $\phasedr$ as expected. Similarly to the threshold gain reduction, the interaction between the two right hand terms determines the lasing frequency stability as discussed in \mbox{Sec.~\ref{sec:ampphase}}.


\section{DERIVING THE RATE EQUATION FOR THE INTENSITY AND PHASE}
\label{sec:aprates}
To further inspect the laser dynamics, the laser rate equations for the intensity and phase must be studied. Considering \mbox{Eqs.~(\ref{eq:fielddefinitions})} \mbox{and~(\ref{eq:field12})}, and following the approach from Lang and Kobayashi \cite{Lang1980},
the laser field equation which considers EOF from both sides of the laser cavity can be written as:
\begin{alignat}{2}
\dot{\mathcal{E}} & =  \left(-i \wref + \Delta G \frac{1 - i\ah}{2}\right) \mathcal{E}(t) + \kr \mathcal{E}(t-\tre) + \kl \mathcal{E}(t+\tle)\nonumber\\
 \Leftrightarrow \frac{d
	\left[ \mathcal{A}(t) e^{-i \wfb t} \right]
}{d t} & \stack{(\ref{eq:feedterm})}{=} e^{-i \wfb t} \left[ \left( -i\wref + \Delta G \frac{1 - i\ah}{2}\right) \mathcal{A}(t)  + \kr \mathcal{A}(t-\tre) e^{i\phasedr} + \kl \mathcal{A}(t+\tre)  e^{-i\phasedl}\right] \nonumber\\
\Leftrightarrow
\dot{\mathcal{A}}(t) & \stack{(\ref{eq:deltaomega})}{=} \stack{(\ref{eq:phased})}{=}\left( i\delta \omega  + \Delta G \frac{1 - i\ah}{2}\right) \mathcal{A}(t) + \kr \mathcal{A}(t-\tre) e^{i\phasedr} + \kl \mathcal{A}(t+\tre)  e^{-i\phasedl}. \label{eq:field_feedback}
\end{alignat}
The last two right-hand terms appear as a result of the imposed feedback conditions, each term to account for feedback on each side of the cavity. In the case without feedback the  lasing frequency become $\wfb = \wref$ and $\ki = 0$, thus recovering the no-feedback field equation~\cite{Henry1982}. 
The slowly varying field amplitude $\mathcal{A}$ can be modeled as \mbox{Eq.~(\ref{eq:amplitude_decomp})}, and therefore the rate equations for the photon number $\I $ and phase $\phi$ can be found using:
\begin{subequations}
	\label{eq:arctanid}
	\setlength{\abovedisplayskip}{0pt}
	\setlength{\belowdisplayskip}{0pt}
	\setlength{\columnsep}{0em}
	\begin{multicols}{2}
		\begin{align}
		\dot{\I } &=  \frac{d[\mathcal{A}\mathcal{A}^{\ast}]}{dt} =  \mathcal{A} \dot{\mathcal{A}^{\ast}} + \mathcal{A}^{\ast} \dot{\mathcal{A}}\label{eq:photonNumber} 
		\end{align}
		\setlength{\abovedisplayskip}{0pt}
		\setlength{\belowdisplayskip}{0pt}
		\begin{alignat}{2} 
		\vspace{1mm}
		\dot{\phi} &= -\frac{1}{\I } \Im\left(\mathcal{A}^{\ast} \dot{\mathcal{A}}\right). 
		\label{eq:phaseE}
		\end{alignat}
	\end{multicols}
\end{subequations}
\noindent Replacing \mbox{Eqs.~(\ref{eq:field_feedback})} \mbox{and~(\ref{eq:amplitude_decomp})} into \mbox{Eq.~(\ref{eq:photonNumber})} yields: 
\begin{alignat}{2}
\dot{\I } 
&  \stack{(\ref{eq:deltaphitl})}{=}  \stack{(\ref{eq:deltaphitr})}{=} \I  \Delta G  
+  \sqsr \sqrt{\I }  \left(e^{-i\Delta \Phitr} +  e^{i\Delta \Phitr}\right) 
+  \sqsl \sqrt{\I }  \left(e^{-i\Delta \Phitl} +  e^{i\Delta \Phitl}\right),
\end{alignat}
and transforming the sum of exponentials into a cosine yields the following expression for the photon rate:
\begin{equation}
\dot{\I } =  \I  \Delta G 
+ 2 \sqsr \sqrt{\I }\cos(\Delta \Phitr)
+ 2 \sqsl \sqrt{\I }\cos(\Delta \Phitl).
\label{eq:finalPhotonRate}
\end{equation}
In the case of the phase, its rate equation comes from replacing \mbox{Eqs.~(\ref{eq:field_feedback})} \mbox{and~(\ref{eq:amplitude_decomp})} into \mbox{Eq.~(\ref{eq:phaseE})}:
\begin{alignat}{2}
\dot{\phi}  \I  
& \stack{(\ref{eq:fielddefinitions})}{=}   - \Im \left[\left(i\delta \omega+ \Delta G \frac{1 - i\ah}{2}\right)\I  
+ \sqsr \sqrt{\I } e^{i\Delta \Phitr}
+ \sqsl \sqrt{\I } e^{i\Delta \Phitl}\right]\nonumber\\
\dot{\phi} &
= \frac{\Delta G}{2}\ah -\delta \omega 
- \frac{\sqsr}{\sqrt{\I }} \sin\left(\Delta \Phitr \right)
- \frac{\sqsl}{\sqrt{\I }} \sin\left(\Delta \Phitl \right).
\label{eq:finalPhaseRate}
\end{alignat}
\mbox{Eqs.~(\ref{eq:finalPhotonRate})} \mbox{and~(\ref{eq:finalPhaseRate})} are the amplitude and phase rate equations for the laser system proposed in this work. These are the starting point to compute the frequency noise PSD, and extract the intrinsic linewidth. 
%
%
%


\section{SMALL-SIGNAL ANALYSIS}
\label{sec:apsmallsignal}
To find the FN PSD, the system shown in \mbox{Eq.~(\ref{eq:rateposta})} is to be solved. This is done using the small-signal analysis proposed in \mbox{Eq.~(\ref{eq:def})}. Assuming a narrow-linewidth laser, \textit{i.e.} a long coherence time with respect to the external cavity lenghts: 
\begin{equation}
\taue < \tcoh,
\label{eq:approxLinearization}
\end{equation}
\noindent the following approximation is valid:
\begin{equation}
\freq \taue \ll  1.
\label{eq:papprox}
\end{equation}
By linearizing the following expressions:
\begin{subequations}
	\begin{alignat}{2}
	\sqrt{\frac{\I (t \pm \tie)}{\I (t) }} \stack{(\ref{eq:approxLinearization})}{\approx}  &\; \sqrt{\frac{\I \pm \dot{\I} \tie }{\I}} 
	\; =  \; \sqrt{ 1 \pm \frac{\dot{\I}}{\I} \tie } 
	\;  \stack{(\ref{eq:pdef})}{\approx}  \; \sqrt{ 1 \pm \frac{\dpd}{\meanI } \tie } 
	\;  \stack{{(\ref{eq:papprox})}}{\approx}  \; 1 \pm \frac{\dpd}{2\meanI } \tie \\
	\phi-\phi(t \pm \tie) \stack{(\ref{eq:approxLinearization})}{\approx} &\;  \phi-\phi \mp \tie  \dot{\phi} \;  \stack{(\ref{eq:phidef})}{=}  \;   \mp i \tie  \Omega' \phi_\Delta ,
	\end{alignat}
	\label{eq:linearepsilonphi}
\end{subequations}
it is possible to rewrite \mbox{Eqs.~(\ref{eq:rateposta})} as:
\begin{subequations}
	\label{eq:carrierlinear_aux}
	\begin{alignat}{2}
	\dpd & \stack{(\ref{eq:def})}{=} \stack{(\ref{eq:linearepsilonphi})}{=} 
	\Delta G \I 
	+ \R  
	+ 2 \kr \I \left( 1 - \frac{\dpd}{2\meanI } \tre  \right)  \cos\left(\dotphip \tre  + \phasedr\right)\nonumber \\
	& \qquad
	+ 2 \kl \I \left( 1 + \frac{\dpd}{2\meanI } \tle  \right)  \cos\left(- \dotphip \tle  - \phasedl\right) 
	+ \FI \label{eq:linearp_aux}\\
	\dotphip &  \stack{(\ref{eq:def})}{=} \stack{(\ref{eq:linearepsilonphi})}{=}  
	\ah \frac{\Delta G}{2} 
	- \delta \omega 
	- \kr \left( 1 - \frac{\dpd}{2\meanI } \tre  \right) \sin\left(\dotphip \tre  + \phasedr \right) \nonumber \\
	&\qquad
	- \kl \left( 1 + \frac{\dpd}{2\meanI } \tle  \right) \sin\left(-\dotphip \tle  - \phasedl\right)  
	+\Fphi \label{eq:linearphi_aux}\\
	\dotnop& \stack{(\ref{eq:def})}{=} \stack{(\ref{eq:linearepsilonphi})}{=}  I - G_{_\mathrm{FB}}\I - \rrec + \FN  \label{eq:carrierdelta}. 
	\end{alignat}
\end{subequations}
Using the following identities:
\begin{subequations}
	\label{eq:parity}
	\setlength{\abovedisplayskip}{-10pt}
	\setlength{\belowdisplayskip}{-10pt}
	\setlength{\columnsep}{0em}
	\begin{multicols}{2}
		
		\begin{align}
		\cos(x) & =  \cos(-x) \label{eq:evencos}
		\end{align}
		
		\begin{alignat}{2}
		\sin(x) & = - \sin(-x), \label{eq:oddsin}
		\end{alignat}
	
	\end{multicols}
\end{subequations}
\noindent the rate equations from  \mbox{Eqs.~(\ref{eq:carrierlinear_aux})} can be rewritten as:
\begin{subequations}
	\label{eq:carrierlinear}
	\begin{alignat}{2}
	\dpd & \stack{(\ref{eq:parity})}{=} 
	 \I  \Delta G 
	+ \R  
	+ 2 \frac{\kr}{\tcav} \I \left( 1 - \frac{\dpd}{2\meanI } \tre  \right)  \cos\left(\dotphip \tre  + \phasedr\right) \nonumber \\
	& \qquad
	+ 2 \kl \I \left( 1 + \frac{\dpd}{2\meanI } \tle  \right)  \cos\left(\dotphip \tle  + \phasedl\right) 
	+ \FI \label{eq:linearp}\\
	\dotphip &   \stack{(\ref{eq:parity})}{=}  
	\ah \frac{\Delta G }{2} 
	- \delta \omega 
	-\frac{\kr}{\tcav} \left( 1 - \frac{\dpd}{2\meanI } \tre  \right) \sin\left(\dotphip \tre  + \phasedr \right) \nonumber \\
	&\qquad
	+\kl \left( 1 + \frac{\dpd}{2\meanI } \tle  \right) \sin\left(\dotphip \tle  + \phasedl\right)  
	+\Fphi. \label{eq:linearphi}
	\end{alignat}
\end{subequations}
\noindent Solving \mbox{Eq.~(\ref{eq:carrierlinear_aux})} requires the steady-state solutions of \mbox{Eq.~(\ref{eq:rateposta})}. Under stationary conditions:
\begin{subequations}
\begin{alignat}{2}
\dot{\I} & = 0 \Rightarrow \I (t) = \I(t \pm \tie ) = \meanI \\
\dot{\phi} & = 0 \Rightarrow \phi(t) = \phi(t \pm \tie ) \\
\dot{N} & = 0 \Rightarrow N(t) = N(t \pm \tie ) = N_{0},
\end{alignat}
\end{subequations}
with which the steady-state equations are:
\begin{subequations}
	 \label{eq:steadystate}
\begin{alignat}{2}
\tau_{\mathrm{ph}}^{-1} &= \go  + 2 \kr \cos \left( \phasedr \right) + 2 \kl \cos \left( \phasedl \right) + \frac{\R}{\meanI}  \\
\delta \omega &= \ah \frac{\go - \tau_{\mathrm{ph}}^{-1}}{2}
- \kr \sin\left( \phasedr \right) + \kl \sin\left( \phasedl \right)
\\
I  &=\go \meanI +\rrecmean,
\end{alignat}
\end{subequations}
where the Langevin noise terms are not includes as their mean value is zero.
\noindent Next, using \mbox{Eq.~(\ref{eq:steadystate})} and the following expansions:
\begin{subequations}
	\label{eq:taylortrig} 
	\begin{multicols}{2}
		\setlength{\abovedisplayskip}{-5pt}
		\setlength{\belowdisplayskip}{0pt}
		\setlength{\columnsep}{0em}
		\begin{alignat}{2}
		\sin(x + \Delta) & \approx \sin(x) + \Delta \cos \left( x\right) 
		\end{alignat}
		\setlength{\abovedisplayskip}{-5pt}
		\setlength{\belowdisplayskip}{0pt}
		\begin{alignat}{2}
		\hspace{-5mm}\cos(x + \Delta)  & \approx \cos(x) - \Delta \sin\left( x \right),
		\end{alignat}
	\end{multicols}
\end{subequations}
\noindent \mbox{Eq.~(\ref{eq:linearp})} can be rewritten as:
\begin{alignat}{2}
	\dpd   \stack{(\ref{eq:def})}{=} \stack{(\ref{eq:steadystate})}{=}  \stack{(\ref{eq:abgammacoefficients})}{\approx} &\,
	\left(\go  +  \again  \nop  -\left\{\go  + 2 \left[\kr \cos \left( \phasedr \right) + \kl \cos \left( \phasedl \right)\right] + \frac{\R}{\meanI}   \right\}\right)
	\left( \meanI  + \pd \right) 
	 \nonumber \\
	& \hspace{-5mm}
	+ 2  \bigg\{ \kr \left(\meanI  + \pd \right) \left[\cos(\phasedr) - \dotphip \tre  \sin\left( \phasedr \right)\right] \nonumber \\
	&- \kr \left(\meanI  + \pd \right) \frac{\dpd}{2 \meanI } \tre  \left[\cos(\phasedr) - \dotphip \tre  \sin\left( \phasedr \right)\right]  \nonumber \\
	& + \kl \left(\meanI  + \pd \right) \left[\cos(\phasedl) - \dotphip \tle  \sin\left( \phasedl \right)\right] \nonumber \\
	& + \kl \left(\meanI  + \pd \right) \frac{\dpd}{2 \meanI } \tle  \left[\cos( \phasedl ) - \dotphip \tle  \sin\left( \phasedl \right)\right] \bigg\} + \R + \FI. \nonumber
\end{alignat}
Simplifying this equation, and neglecting the quadratic terms yields:
\begin{subequations}
	\begin{alignat}{2}
	\dpd  &  \stack{(\ref{eq:abgammacoefficients})}{\approx} \, \again  \nop \meanI - \gammai  \pd  - \dotphip 2  \mathsf{K}_{s} \meanI + \dpd (\kcl - \kcr)    + \FI . \label{eq:prate}
	\end{alignat}
In a similar way, \mbox{Eqs.~(\ref{eq:linearphi})} can be rewritten using \mbox{Eqs.~(\ref{eq:steadystate})} and \mbox{(\ref{eq:abgammacoefficients})}: 
\begin{alignat}{2}
	\dotphip & \approx \, 
	\frac{\ah}{2}\left(\go  +  \again  \nop - \tau_{\mathrm{ph}}^{-1}\right)  
	-\left[ \ah \frac{\go - \tau_{\mathrm{ph}}^{-1}}{2}
	- \kr \sin\left( \phasedr \right) + \kl \sin\left( \phasedl \right) \right] \nonumber\\
	& \quad - \kr \left(1 - \frac{\dpd}{2\meanI } \tre \right) 
	\left[\sin\left( \phasedr \right)+ \dotphip \tre  \cos \left( \phasedr \right)\right]\nonumber\\
	& \quad + \kl \left(1 + \frac{\dpd}{2\meanI } \tle \right) 
	\left[\sin\left( \phasedl \right)+ \dotphip \tle  \cos \left( \phasedl \right)\right]
	+ \Fphi \nonumber\\
	\Leftrightarrow 2 \dotphip & \stack{(\ref{eq:acoefficients})}{=} \stack{(\ref{eq:bcoefficients})}{=} \,  
	\ah  \again  \nop 
	+  2 (\kcl - \kcr)  \dotphip 
	+ \frac{\dpd}{\meanI } \mathsf{K}_{s}
	+ 2\Fphi. \label{eq:phirate}
\end{alignat}
Finally, \mbox{Eq.~(\ref{eq:carrierdelta})} can be rewritten as:
\begin{alignat}{3}
	\dotnop \stack{(\ref{eq:def})}{=} \stack{(\ref{eq:steadystate})}{=} & \stack{(\ref{eq:abgammacoefficients})}{\approx} \, 
	&& \, \go \meanI   - \left(\go + \again  \nop \right) \meanI  -  \left(\go  + \again  \nop \right)\pd - \rrecN \nop  +\FN \nonumber \\
	\Leftrightarrow	\dotnop & \stack{(\ref{eq:gammaN})}{=}\, && - \gamman \nop  - \go  \pd +\FN. \label{eq:carrierrate}
\end{alignat}
\end{subequations}
The linearized rate equations of the laser under study are thus \mbox{Eqs.~(\ref{eq:prate})}, (\ref{eq:phirate})  \mbox{and~(\ref{eq:carrierrate})}, from which the power spectral density, and subsequently the linewidth, can be computed.


\section{POWER SPECTRAL DENSITY}
\label{sec:appsd}
Obtaining the FN PSD of the system under study requires calculating $\hatphip$.
To begin with, $\hatnop$ is extracted from \mbox{Eq.~(\ref{eq:fourier_n})}:
\begin{alignat}{2}
\left(i \freq + \gamman\right)\hatnop & =  - \go  \hatpd + \hatFN \quad
\Leftrightarrow \quad  \hatnop = \, \frac{ \hatFN  - \go  \hatpd}{i \freq + \gamman}, \label{eq:hatn}
\end{alignat}
Next, replacing \mbox{Eq.~(\ref{eq:hatn})} into \mbox{Eq.~(\ref{eq:fourier_p})} yields the expression for $\hatpd$:
\begin{alignat}{2}
\left[i \freq \mathsf{K}_{c} + \gammai \right]\hatpd 
& = \again \frac{ \hatFN - \go  \hatpd }{i \freq + \gamman} \meanI  
- i 2 \freq  \meanI \mathsf{K}_{s} \hatphip
+ \hatFI \nonumber \\
\Leftrightarrow	\frac{\Aphi}{i \freq + \gamman} \hatpd & 
= \,\again  \meanI    \frac{\hatFN }{i \freq + \gamman}  
- i 2 \freq  \meanI \mathsf{K}_{s} \hatphip 
+ \hatFI \nonumber \\
\Leftrightarrow\hatpd& \stack{(\ref{eq:aphimoni})}{=} \frac{\,\again  \meanI \hatFN
	- i 2  \meanI  \freq \mathsf{K}_{s} \left(i \freq + \gamman \right) \hatphip 
	+\left(i \freq + \gamman \right)  \hatFI}{\Aphi}. \label{eq:hatp}
\end{alignat}
Finally, inserting \mbox{Eq.~(\ref{eq:hatn})} \mbox{and~(\ref{eq:hatp})} into \mbox{Eq.~(\ref{eq:fourier_phi})}
and grouping the terms with $\hatphip$, $\hatFI$, $\hatFphi$ and $\hatFN $ yields:
\begin{alignat}{2}
2 i \freq \deltam \, \hatphip	&  \stack{(\ref{eq:acoefdef})}{=} \left[ -\ah  \GnG  \meanI 
+ \ah \Aphi
+\, i \freq \mathsf{K}_{s} \left(i \freq + \gamman\right)\right] \frac{ \again }{i \freq + \gamman} \, \hatFN  + 2 \AI   \, \hatFI  + 2 \, \Aphi \hatFphi \nonumber\\
& \stack{(\ref{eq:ai})}{=}\left( \ah  \Aphi + 2 \meanI \AI  \right) \frac{ \again }{i \freq + \gamman} \, \hatFN  +  \AI   \, \hatFI  + \, 2 \Aphi \hatFphi \nonumber\\
\Leftrightarrow \hatphip & \stack{(\ref{eq:an})}{=}\frac{
	\AN  \, \hatFN  +  \AI   \, \hatFI  + \, \Aphi \hatFphi}{i \freq \deltam} . \label{eq:phihatpretty_app}
\end{alignat}
With \mbox{Eq.~(\ref{eq:phihatpretty_app})} it is possible to calculate an expression for \mbox{Eq.~(\ref{eq:PSDfromphi})}:
\begin{alignat}{2}
2 \pi^2 |\deltam |^2 \PSD &  \stack{(\ref{eq:PSDfromphi})}{=} \freq^2 |\deltam |^2 \left\langle  \hatphip(\freq) \hatphip^{\ast}(\freq)  \right\rangle  \nonumber\\
& \stack{(\ref{eq:phihatpretty_app})}{=} \freq^2 |\deltam |^2 \left\langle  \frac{\AN  \, \hatFN  +  \AI   \, \hatFI  + \, \Aphi \hatFphi}{i \freq \deltam}
\left(\frac{\AN  \, \hatFN  +  \AI   \, \hatFI  + \, \Aphi \hatFphi}{i \freq \deltam}\right)^{\displaystyle *}  \right\rangle \nonumber\\
&=\left\langle  \left(\AN  \, \hatFN  +  \AI   \, \hatFI  + \, \Aphi \hatFphi \right)
\left(\AN  \, \hatFN  +  \AI   \, \hatFI  + \, \Aphi \hatFphi \right)^{\displaystyle *}  \right\rangle 
 \label{eq:auxaveragephi}
\end{alignat}
It can be seen from \mbox{Eq.~(\ref{eq:acoefdef})} that the coefficients $A_{i}$ are independent of time, and assuming an ergodic process they can be taken out of the average in \mbox{Eq.~(\ref{eq:auxaveragephi})}, obtaining:
\begin{alignat}{2}
2 \pi^2 |\deltam |^2 \PSD & = 
|\AN |^2 \left\langle \hatFN  \, \hatFN ^{\displaystyle *} \right\rangle
+ \AI  \AN ^{\displaystyle *} \left\langle   \hatFI  \hatFN ^{\displaystyle *} \right\rangle 
+ \Aphi \AN ^{\displaystyle *} \left\langle\,  \hatFphi\hatFN ^{\displaystyle *} \right\rangle \nonumber\\
& + \AN  A^{\displaystyle *}_{\I}  \left\langle \hatFN   \hatFI^{\displaystyle *} \right\rangle 
+ |\AI |^2  \left\langle  \hatFI \hatFI^{\displaystyle *} \right\rangle 
\; \, + \Aphi A^{\displaystyle *}_{\I} \left\langle  \hatFphi  \hatFI^{\displaystyle *} \right\rangle \nonumber\\
&+  \AN  A^{\displaystyle *}_{\phi} \left\langle \hatFN   \hatFphi^{\displaystyle *} \right\rangle
+ \AI  A^{\displaystyle *}_{\phi} \left\langle   \, \hatFI \hatFphi^{\displaystyle *} \right\rangle
+ |\Aphi|^2 \left\langle  \hatFphi \hatFphi^{\displaystyle *}  \right\rangle, 
\end{alignat}
which can be rewritten as:
\begin{subequations}
	\begin{alignat}{2}
	\pi^2 |\deltam |^2 \PSD & \stack{(\ref{eq:fourierdiffusion})}{=} 
	|\AN |^2 \dnn
	+ |\AI |^2 \dii
	+ \,|\Aphi|^2 \dpp  + \left(\AI  \AN ^{\displaystyle *} +\AN  A^{\displaystyle *}_{\I} \right) \din.
	\label{eq:avphiphiconjdiff}
	\end{alignat}
\end{subequations}
\noindent  The explicit formula for $\PSD$ can be obtained from 
\mbox{Eq.~(\ref{eq:avphiphiconjdiff})}, which
requires calculating $|\AN |^2$, $|\AI |^2$, $|\Aphi|^2$ and  $\left(\AI  \AN ^{\displaystyle *} +\AN  A^{\displaystyle *}_{\I} \right)$. 
For this purpose, it is helpful to separate the real and imaginary parts of these coefficients. Expanding $\AN$ yields: 
\begin{subequations}
	\begin{alignat}{2}
	2 \AN  & \stack{(\ref{eq:an})}{=}\stack{(\ref{eq:lambdasmoni})}{=}
	\gammai \ah \again 
	+ i \freq  \again  \mathsf{F}_{2} \label{eq:anrealim}\\
	\Leftrightarrow 4 |\AN |^2 & = \left(\gammai \ah \again \right)^2 + \freq^2 \again^2 \mathsf{F}_{2}^2 \, . \label{eq:ansq}
	\end{alignat}
\end{subequations}
In the case of $\Aphi$, the same approach yields:
\begin{subequations}
	\begin{alignat}{2}
	\Aphi & \stack{(\ref{eq:aphimoni})}{=}
	\left(i \freq + \gamman\right) \left(i \freq \mathsf{K}_{c}  + \gammai\right)
	+ \GnG  \meanI   \nonumber\\
	& = - \freq^2\mathsf{K}_{c}  + \gamman \gammai + \GnG  \meanI  + i \freq \left(\gammai + \mathsf{K}_{c}  \gamman\right) \label{eq:aphirealim} \\
	\Leftrightarrow &	|\Aphi|^2  = \left(
	- \freq^2 
	\mathsf{K}_{c} 
	+ \gamman \gammai
	+ \GnG   \meanI 
	\right)^2 + \freq^2 \left(\gammai + \mathsf{K}_{c}  \gamman\right)^2 \nonumber \\
	& \stack{(\ref{eq:upsilon})}{=} \freq^4 \mathsf{K}_{c} ^2 + \freq^{2} \mathsf{F}_{0}
	+ \left(\gamman \gammai + \GnG  \meanI \right)^2,  \label{eq:aphisq} \vspace{-3mm} 
	\end{alignat} 
\end{subequations}
\noindent and in the case for $\AI$: 
\begin{subequations}
	\begin{alignat}{2}
	2  \AI  & \stack{(\ref{eq:ai})}{=} 
	- \ah \GnG
	- \freq^2 \frac{\mathsf{K}_{s} }{ \meanI } 
	+ i \freq \gamman \frac{\mathsf{K}_{s} }{ \meanI }\label{eq:airealim}\\
	\Leftrightarrow
	4 \meanI ^2 |\AI |^2 & =  \left(-\ah \GnG  \meanI 
	- \freq^2 \mathsf{K}_{s}  \right)^2
	+ \freq^2 \left(\mathsf{K}_{s} \gamman\right)^2 \nonumber\\
	& =  \freq^{4} \mathsf{K}_{s}^2 + \freq^{2} \left[\left(\mathsf{K}_{s} \gamman \right)^2 + 2 \ah \GnG \mathsf{K}_{s}  \meanI  
	\right]  +\left(\ah \GnG \meanI \right)^2. \label{eq:aisq}
	\end{alignat}
\end{subequations}
Next, the following identity is useful for computing $ \left(\AI  \AN ^{\displaystyle *} +\AN  A^{\displaystyle *}_{\I} \right)$:
\begin{equation}
XY^{\displaystyle *} +X^{\displaystyle *} Y = (a+ib)(c-id)+(a-ib)(c+id) = 2 (a c + b d), 
\end{equation}
with which:
\begin{alignat}{2}
4 \left(\AI  \AN ^{\displaystyle *} +\AN  A^{\displaystyle *}_{\I} \right)   \stack{(\ref{eq:anrealim})}{=} \stack{(\ref{eq:airealim})}{=} &   2 \left\{
\left[
- \ah \GnG  
- \freq^2  \mathsf{K}_{s} \big/ \meanI 
\right]
\gammai \ah \again  \right. \nonumber \\
& \left. +\freq^2 \gamman \left( \mathsf{K}_{s} \big/ \meanI\right)
\again  \left(\mathsf{K}_{c} \ah + \mathsf{K}_{s} \right)
\right\}\nonumber\\
\Leftrightarrow 2 \left(\AI  \AN ^{\displaystyle *} +\AN  A^{\displaystyle *}_{\I} \right) 
= 
\freq^2  \frac{\mathsf{K}_{s}}{\meanI }  & \again  \left[\gamman  \left(\mathsf{K}_{c} \ah + \mathsf{K}_{s} \right) - \gammai \ah \right]
- \ah^2 \again^2  \go  \gammai
. \label{eq:aian}
\end{alignat}
Using \mbox{Eq.~(\ref{eq:ansq})}, (\ref{eq:aphisq}), (\ref{eq:aisq}), (\ref{eq:aian}) \mbox{and~(\ref{eq:deltamoni_aux})}, an expression for  $\PSD$ can be found:
\begin{alignat}{2}
4 \pi^2 |\deltam |^2 \PSD  \stack{(\ref{eq:avphiphiconjdiff})}{=} \stack{(\ref{eq:lambdasmoni})}{=} &
\left[\left(\gammai \ah \again \right)^2 + \freq^2  \again^2 \mathsf{F}_{2}^2 \right] \dnn \nonumber \\
& + 
\left[
\freq^4  \mathsf{K}_{c}^2 + \freq^{2} \mathsf{F}_{0} + \left(\gamman \gammai + \GnG  \meanI \right)^2 
\right]
4\dpp \nonumber\\
& + \left\{
\freq^{2} \left[\left( \gamman \frac{\mathsf{K}_{s} }{\meanI}\right)^2 + 2 \ah \GnG  \frac{\mathsf{K}_{s} }{\meanI}
\right]  + \freq^{4} \left(\frac{\mathsf{K}_{s} }{\meanI}\right)^2 +\left[\ah \GnG \right]^2
\right\} \dii   \nonumber\\
& - 2 \left[ - \ah^2 \again^2  \go  \gammai +\freq^2 \frac{\mathsf{K}_{s} }{ \meanI } \again  \left(\gamman  \mathsf{F}_{2} - \gammai \ah \right)  \right] \dii \nonumber \\
\Leftrightarrow 2 \pi^2 \PSD & \stack{(\ref{eq:lambdasmoni})}{=} \frac{\Lambda_{4} \freq^4 + \Lambda_{2} \freq^2 + \Lambda_{0}}{2 |\deltam|^2 }. \label{eq:deltaphiomegamoni_app}
\end{alignat}
\noindent Furthermore, using the following definitions:
\begin{subequations}
	\begin{alignat}{2}
	\deltafour& \equiv \mathsf{F}_{1}^2 \\
	\deltatwo & \equiv \mathsf{K}_{c}^2 \gammai^2
	+\gammansq \mathsf{F}_{1}^2
	- 2 \mathsf{F}_{1}  \GnG  \meanI  \mathsf{F}_3 \\
	\deltazero &\equiv \left(\GnG  \meanI   \mathsf{F}_{3} 
	+ \mathsf{K}_{c} \gammai \gamman \right)^2\!, 	\label{eq:defdeltasmoni}
	\end{alignat}
\end{subequations}
 the expression for $\deltam$ from \mbox{Eq.~(\ref{eq:deltamoni_aux})} can be rewritten as:
\begin{subequations}
	\label{eq:deltacomp}
	\begin{alignat}{2}
	\deltam 
	& =  \GnG \meanI    \mathsf{F}_{3} - \freq^2 \mathsf{F}_{1}
	+ i \freq \left(\mathsf{K}_{c} \gammai 
	+\gamman \mathsf{F}_{1}\right) + \mathsf{K}_{c} \gammai\gamman  \label{eq:deltamoniexpanded}\\
	\Leftrightarrow |\deltam|^2 	& = \freq^2 \left(\mathsf{K}_{c} \gammai 
	+\gamman \mathsf{F}_{1}\right)^2 
	+ 
	\left(\GnG  \meanI   \mathsf{F}_{3} - \freq^2 \mathsf{F}_{1} + \mathsf{K}_{c} \gammai\gamman\right)^2 \nonumber\\
	& \stack{(\ref{eq:defdeltasmoni})}{=} \freq^4  \deltafour
	+ \freq^2 \left[\left(\mathsf{K}_{c}  \gammai
	+\gamman \mathsf{F}_{1}\right)^2
	- 2 \mathsf{F}_{1} \GnG  \meanI  \mathsf{F}_3
	- 2 \mathsf{F}_{1} \mathsf{K}_{c} \gammai\gamman 
	\right] + \deltazero \nonumber\\
	& \stack{(\ref{eq:defdeltasmoni})}{=} \freq^4 \deltafour + \freq^2 \deltatwo+ \deltazero, \label{eq:deltasmoni}
	\end{alignat}
\end{subequations}
with which the expression of the FN PSD becomes:
\begin{equation}
4 \pi^2 \PSD  \stack{(\ref{eq:lambdasmoni})}{=} \frac{\Lambda_{4} \freq^4 + \Lambda_{2} \freq^2 + \Lambda_{0}}{\deltafour \freq^4  + \deltatwo \freq^2  + \deltazero }. \label{eq:deltaphiomegamoni_app}
\end{equation}


\section{EXPRESSION FOR THE INTRINSIC LINEWIDTH}
\label{sec:aplinewidth}
The laser intrinsic linewidth can be found from \mbox{Eq.~(\ref{eq:psdlinewidth})}.
Defining:
\begin{alignat}{2}
\beta_{_{Ag}} \equiv  \frac{ \gamman \gammai } {\GnG  \meanI }  \qquad ; \qquad
\delta_{_{Ag}} \equiv \frac{\gammai}{\go }  \qquad ; \qquad
\Delta f_{0} = \frac{\R }{4 \pi \meanI },
\end{alignat}
where, following Ref. [\citenum{Agrawal1984}]: 
\begin{equation}
\delta_{Ag} \simeq 0 \simeq \beta_{_{Ag}} \label{eq:assumptiondeltabeta},
\end{equation}
as $\delta_{Ag}< 10^{-2}$, which accounts for shot noise in the generation and recombination of minority carriers, and $\beta_{_{Ag}}$ is inversely proportional to the laser power which above threshold becomes negligible. %
Starting from \mbox{Eq.~(\ref{eq:deltaphiomegamoni_app})} and setting $\freq = 0$ as required by \mbox{Eq.~(\ref{eq:psdlinewidth})}:
\begin{alignat}{2}
4 \pi \Delta f & = \Lambda_{0} \big/ \deltazero \nonumber\\
& \stack{(\ref{eq:defdeltasmoni})}{=} \stack{(\ref{eq:lambdasmoni})}{=}
\frac{
	\left(\gammai \ah \again \right)^2 \left[\R  \meanI \left(1 + \frac{\left(\go\right)^2}{\gammai^2} + \frac{2 \go }{\gammai}\right) + \rrec\right]
	+ \left(\gamman \gammai + \GnG  \meanI \right)^2  \frac{\R }{\meanI }
}
{\left(\GnG  \meanI   \mathsf{F}_3
	+ \mathsf{K}_{c} \gammai \gamman \right)^2}  \nonumber \\
\Leftrightarrow \frac{\Delta f}{\Delta f_0}& =
\frac{
	\left( \frac{\ah}{\go } \right)^2 
	\left(\gammai^2 + \go^2 + 2 \go \gammai + \frac{\gammai^2 \rrec }{\R \meanI } \right)
	+ \left( \beta_{_{Ag}} +1\right)^2 
}
{\left(\mathsf{K}_{c} \beta_{_{Ag}} 
	+ \mathsf{F}_3 \right)^2} \nonumber \\
& =
\frac{\left( \beta_{_{Ag}} +1\right)^2  + 
	\ah^2 
	\left[1 + 2 \delta_{Ag} + \delta_{Ag}^2 \left(1 + \frac{ \rrec }{\R \meanI }\right) \right]
}
{\left( \mathsf{K}_{c} \beta_{_{Ag}} 
	+  \mathsf{F}_3 \right)^2} \nonumber
\end{alignat}
\begin{alignat}{2}
\stack{(\ref{eq:assumptiondeltabeta})}{\Leftrightarrow} \frac{\Delta f}{\Delta f_{0} \left( 1 + \ah^2 \right)} & \simeq 
\mathsf{F}_3^{-2}  \nonumber \\
& \stack{(\ref{eq:lambdasmoni})}{=} \stack{(\ref{eq:abgammacoefficients})}{=} 
\left\{1+ \kr \tre  \left[\cos\left( \phasedr \right) - \ah \sin\left( \phasedr\right) \right] - \kl \tle  \left[\cos\left( \phasedl \right) + \ah \sin\left( \phasedl\right) \right]\right\}^{-2}
\nonumber\\
& \stack{(\ref{eq:sinarctan})}{=} \stack{(\ref{eq:cosarctan})}{=}
 \left[1+ \gamma_{_{\mathrm{H}}} \kr \tre  \cos \left( \phasedr + \theta_{_{\mathrm{H}}}\right) - \gamma_{_{\mathrm{H}}} \kl \tle  \cos \left( \phasedl - \theta_{_{\mathrm{H}}} \right)\right]^{-2}. \label{eq:correctingAgrawal_app}
\end{alignat}
The presence of EOF from both sides of the laser cavity results in two terms in the linewidth expression, one for each side, as was seen in the threshold gain reduction and lasing frequency shift due to feedback. This result is discussed in \mbox{Sec.~\ref{sec:linepsd}}.

\pagebreak

\section{Suplementary images: Tolerances}

\begin{figure}[h]
	\begin{center}
		\includegraphics[width=0.75\linewidth]{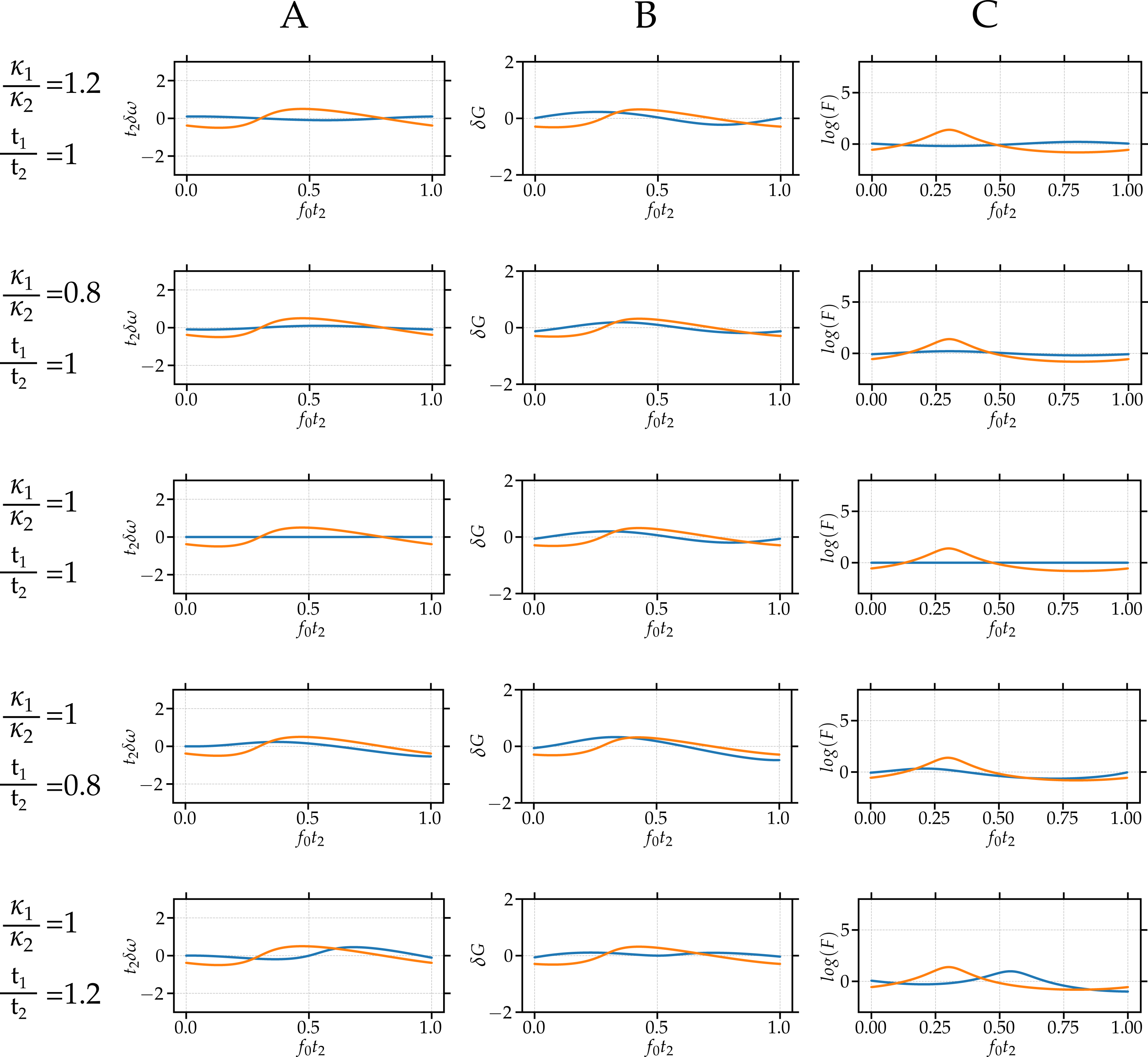}
		\caption{Simulations for $C=0.5$ under \mbox{condition (\ref{eq:phim_condition})} for a $\pm 20\%$ variation of \mbox{conditions (\ref{eq:kappa_assumption})} \mbox{and (\ref{eq:text_assumption})}. Full solution shown in blue, single feedback case shown in orange. Column A shows the lasing frequency shift results. Column B shows the threshold gain shift. Column C shows the intrinsic linewidth variations.}
		\label{fig:tolcless1}
	\end{center}
\end{figure}

\begin{figure}[h]
	\begin{center}
		\includegraphics[width=0.75\linewidth]{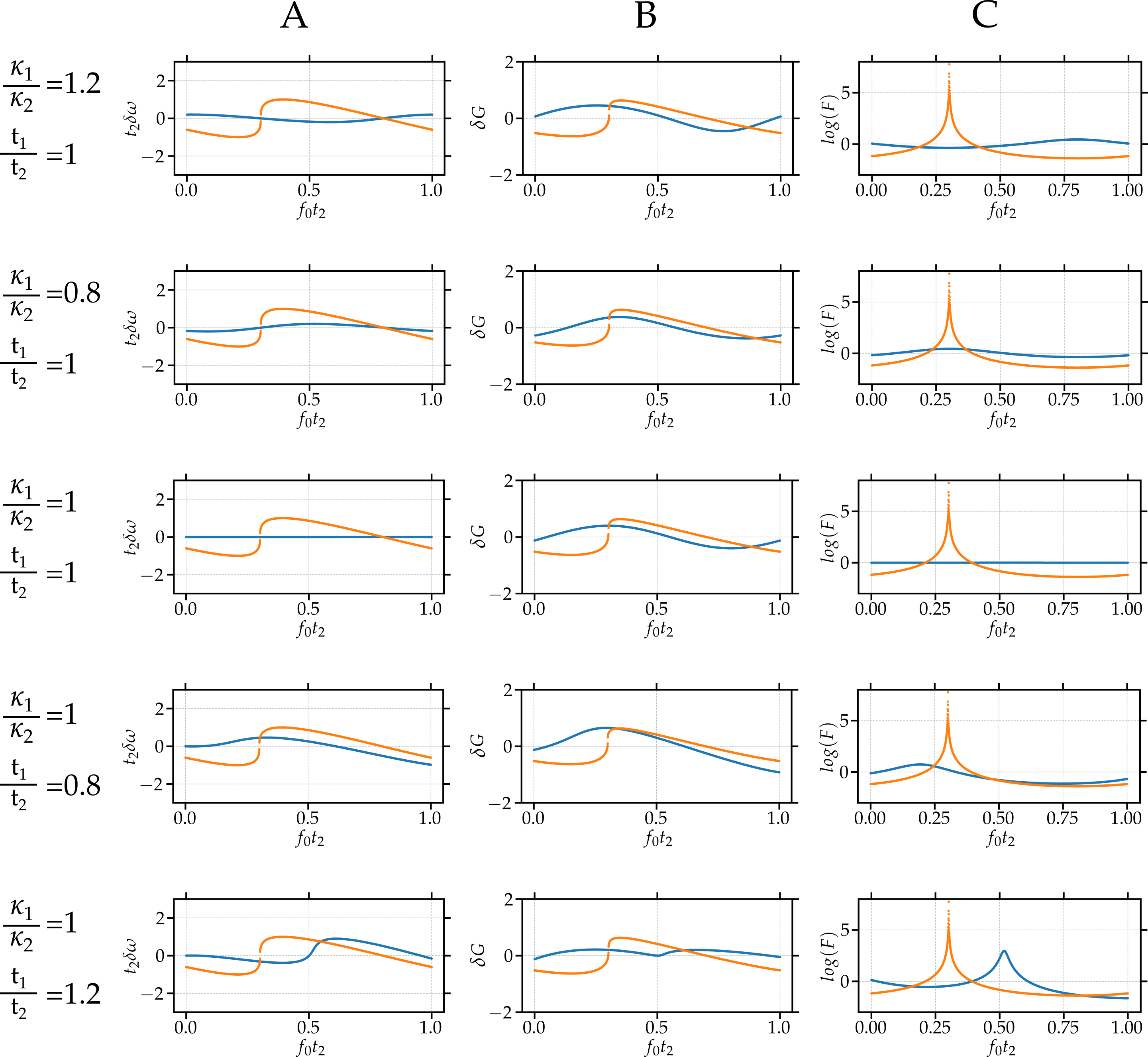}
		\caption{Simulations for $C=1$ under \mbox{condition (\ref{eq:phim_condition})} for a $\pm 20\%$ variation of \mbox{conditions (\ref{eq:kappa_assumption})} \mbox{and (\ref{eq:text_assumption})}. Full solution shown in blue, single feedback case shown in orange. Column A shows the lasing frequency shift results. Column B shows the threshold gain shift. Column C shows the intrinsic linewidth variations.}
		\label{fig:tolcequal1}
	\end{center}
\end{figure}

\begin{figure}[h]
	\begin{center}
		\includegraphics[width=0.75\linewidth]{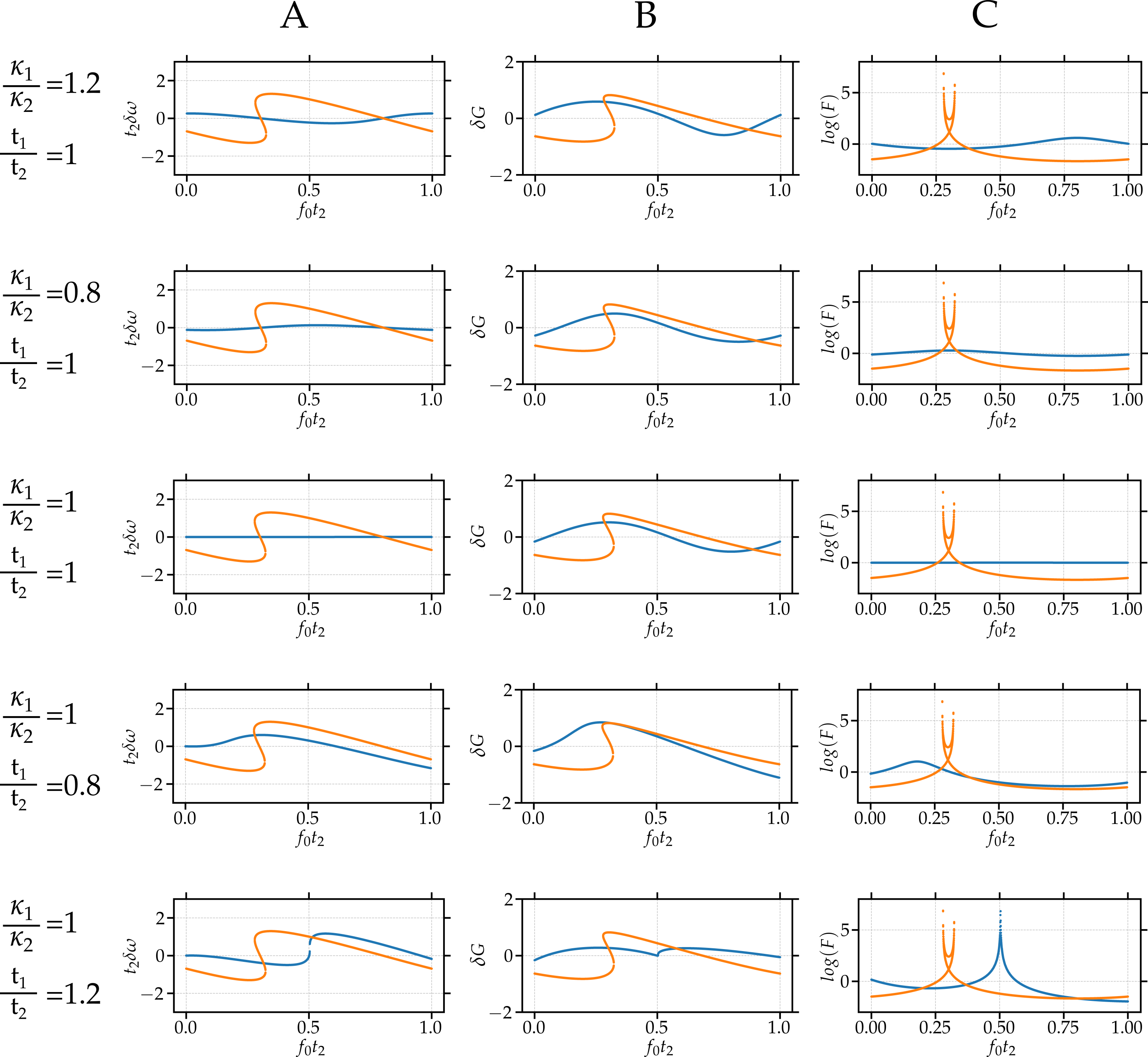}
		\caption{Simulations for $C=1.3$ under \mbox{condition (\ref{eq:phim_condition})} for a $\pm 20\%$ variation of \mbox{conditions (\ref{eq:kappa_assumption})} \mbox{and (\ref{eq:text_assumption})}. Full solution shown in blue, single feedback case shown in orange. Column A shows the lasing frequency shift results. Column B shows the threshold gain shift. Column C shows the intrinsic linewidth variations.}
		\label{fig:tolcmore1}
	\end{center}
\end{figure}
%
%

\pagebreak
 \clearpage 

\bibliographystyle{bibliostyle_perso_2013_01_10}
\bibliography{biblio_MDPI}

\end{document}